\documentclass[10pt,journal,compsoc]{./IEEEtran}

\ifCLASSOPTIONcompsoc
  \usepackage[nocompress]{cite}
\else
  \usepackage{cite}
\fi

\newcommand{\tablestyle}[2]{\setlength{\tabcolsep}{#1}\renewcommand{\arraystretch}{#2}\centering\footnotesize}
\newlength\savewidth
\usepackage[table]{xcolor}

\ifCLASSINFOpdf
  \usepackage[pdftex]{graphicx}
  \graphicspath{{./pics/}}
  \DeclareGraphicsExtensions{.pdf,.jpeg,.png}
\else
\fi

\usepackage{amsmath}
\usepackage{subfigure}

\usepackage{algorithmic}
\usepackage[linesnumbered,ruled]{algorithm2e}

\usepackage{url}

\usepackage{epsfig}
\usepackage{amssymb}
\usepackage[pagebackref=true,breaklinks=true,letterpaper=true,colorlinks,bookmarks=false]{hyperref}

\usepackage{xcolor}
\usepackage{xspace}
\usepackage{multirow}
\usepackage{booktabs}
\usepackage{placeins}
\usepackage{amssymb}
\usepackage{pifont}
\usepackage{bm}
\usepackage{threeparttable}
\usepackage{makecell}
\usepackage{verbatim}
\usepackage{balance}

\usepackage{multirow}
\usepackage{times}
\usepackage{epsfig}
\usepackage{amsmath} 
\usepackage{makecell}

\usepackage{amsmath,amsfonts,bm}

\def\eqref#1{equation~\ref{#1}}

\def\1{\bm{1}}

\newcommand{\cG}{\mathcal{G}}

\def\vg{{\bm{g}}}

\newcommand{\bc}{\mathbf{c}}

\newcommand{\bG}{\mathbf{G}}

\newcommand{\bJ}{\mathbf{J}}

\newcommand{\bn}{\mathbf{n}}\newcommand{\bN}{\mathbf{N}}
\newcommand{\bo}{\mathbf{o}}
\newcommand{\bp}{\mathbf{p}}

\newcommand{\br}{\mathbf{r}}\newcommand{\bR}{\mathbf{R}}
\newcommand{\bS}{\mathbf{S}}
\newcommand{\bt}{\mathbf{t}}\newcommand{\bT}{\mathbf{T}}
\newcommand{\bu}{\mathbf{u}}
\newcommand{\bv}{\mathbf{v}}\newcommand{\bV}{\mathbf{V}}
\newcommand{\bw}{\mathbf{w}}
\newcommand{\bx}{\mathbf{x}}

\DeclareMathAlphabet{\mathsfit}{\encodingdefault}{\sfdefault}{m}{sl}
\SetMathAlphabet{\mathsfit}{bold}{\encodingdefault}{\sfdefault}{bx}{n}

\newcommand{\y}{\bm{y}}
\newcommand{\z}{\mathbf{z}}
\newcommand{\gz}{\bm{z}}

\newcommand{\Vv}{\bm{v}}

\newcommand\myldots{\ifmmode\ldots\else\makebox[0.5em][c]{.\hfil.\hfil.}\thinspace\fi}
\newcommand\mycdots{\ifmmode\cdots\else\makebox[0.5em][c]{.\hfil.\hfil.}\thinspace\fi}

\usepackage{ragged2e}
\definecolor{yellow}{rgb}{1, 1, 0.7}
\definecolor{orange}{rgb}{1, 0.85, 0.7}
\definecolor{tablered}{rgb}{1, 0.7, 0.7}
\definecolor{red}{rgb}{1, 0, 0}
\usepackage{colortbl}
\newcommand{\best}{\cellcolor{yellow}}
\newcommand{\sbest}{\cellcolor{white}}

\DeclareMathOperator{\SE}{SE}

\DeclareMathOperator{\SO}{SO}

\begin{document}

\title{Video4DGen: Enhancing Video and 4D Generation through Mutual Optimization}

\author{Yikai Wang, Guangce Liu, Xinzhou Wang, Zilong Chen, Jiafang Li, Xin Liang,\\  Fuchun Sun,~\IEEEmembership{Fellow,~IEEE,} Jun Zhu,~\IEEEmembership{Fellow,~IEEE}}

\markboth{IEEE Transactions on Pattern Analysis and Machine Intelligence}
{Shell \MakeLowercase{\textit{et al.}}: Bare Demo of IEEEtran.cls for Computer Society Journals}

\IEEEtitleabstractindextext{
\begin{abstract}
The advancement of 4D (\emph{i.e.}, sequential 3D) generation  opens up new possibilities for lifelike experiences in various applications, where users can explore dynamic objects or characters from any viewpoint. Meanwhile, video generative models are receiving particular attention given their ability to produce  realistic and imaginative frames. These models are also observed to exhibit strong 3D consistency,  indicating the potential to act as world simulators.  In this work, we present Video4DGen, a novel framework that excels in  generating 4D representations from single or  multiple generated videos as well as generating 4D-guided videos. This framework is pivotal for creating high-fidelity virtual contents that maintain both spatial and temporal coherence. The  4D  outputs generated by Video4DGen are represented using our proposed {Dynamic Gaussian Surfels} (DGS), which optimizes time-varying warping functions to transform Gaussian surfels (surface elements) from a static state to a dynamically warped state. We design warped-state geometric regularization and refinements on Gaussian surfels, to preserve the  structural integrity and  fine-grained appearance details, respectively. Additionally, in order to perform 4D generation from multiple videos and effectively capture representation across spatial, temporal, and pose dimensions, we design multi-video alignment, root pose optimization, and pose-guided frame sampling strategies. The leveraging of  continuous warping fields also enables a precise depiction of pose, motion, and deformation over per-video frames. Further, to improve the overall fidelity from the observation of all camera poses, Video4DGen  performs novel-view video generation  guided by the  4D content, with the proposed confidence-filtered DGS to enhance the quality of generated sequences. In summary, Video4DGen yields dynamic 4D generation with the ability to handle different subject movements, while preserving details in both geometry and appearance. The framework also generates 4D-guided videos with high spatial and temporal coherence.  With the ability of 4D and video generation, Video4DGen offers a powerful tool for applications in virtual reality, animation, and beyond. See \href{https://github.com/yikaiw/vidu4d}{code} and \href{https://video4dgen.github.io}{project page} for more details. 
\end{abstract}

}

\maketitle

\IEEEdisplaynontitleabstractindextext

\IEEEpeerreviewmaketitle

\section{Introduction}
\IEEEPARstart{T}{he} increasing demand for engaging and interactive digital environments has elevated the importance of generating lifelike, dynamic multimodal content, such as 4D and videos,  holding great promise for various applications. This dynamic multimodal generation process often involves capturing not just spatial and visual details, but also the temporal dynamics of motion, making it crucial to ensure that objects or scenes move fluidly and consistently across multiple frames and viewpoints. 

Recently, video generative models have garnered attention for their remarkable capability to craft immersive and lifelike frames~\cite{videoworldsimulators2024,bao2024vidu}. These models produce visually stunning content while also exhibiting strong 3D consistency~\cite{chen2024v3d,voleti2024sv3d}, largely increasing their potential to simulate realistic environments. Parallel to these developments, 4D reconstruction has made great strides~\cite{pumarola2021d, TiNeuVox, DBLP:journals/tog/ParkSHBBGMS21, yang2023deformable3dgs, wu20234dgaussians}, which involves capturing and rendering detailed spatial and temporal information. When integrated with generative video technologies, this technique potentially enables the creation of models that capture static scenes and dynamic sequences over time. This synthesis provides a  holistic representation of reality,  crucial for applications such as virtual reality, scientific visualization, and embodied artificial intelligence.

\begin{figure}[!ht]
\centering  \vskip-0.05in
\subfigure[Prompt: A portrait captures the dignified presence of an orange cat with striking blue eyes. The cat wears a single pearl earring. Her head tilts in contemplation, reminiscent of a Dutch cap.
]{\includegraphics[width=1\linewidth]{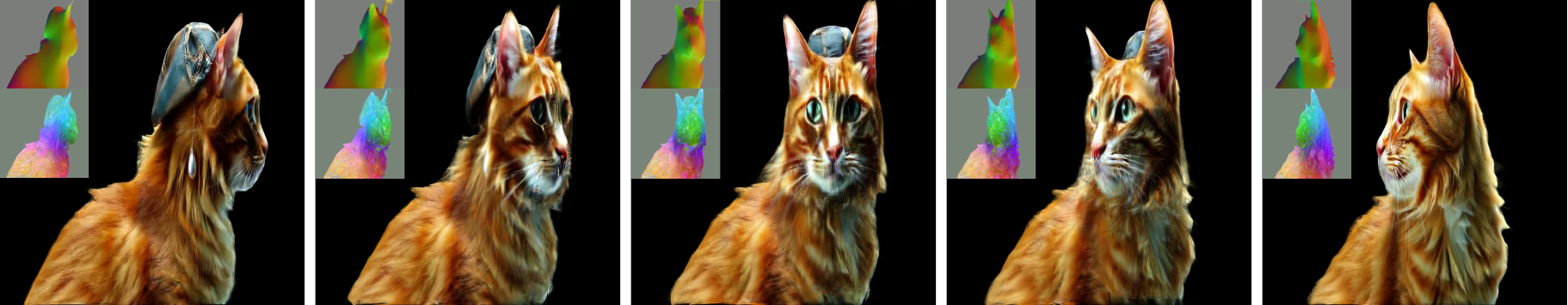}\label{fig:img_cat}}\\ \vskip-0.001in
\subfigure[Prompt: A dragon with its hair blown by a strong wind. Devil enters the  soul with ethereal landscapes.]{\includegraphics[width=1\linewidth]{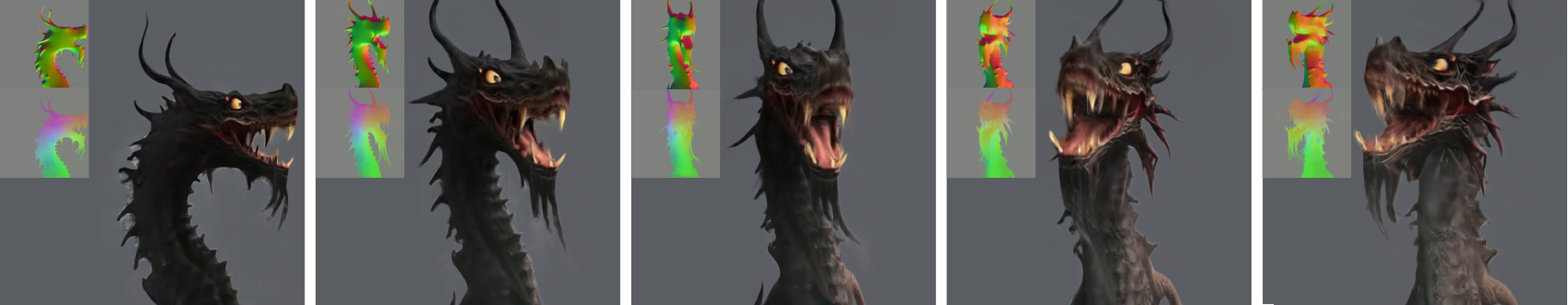}\label{fig:img_dragon}} \vskip-0.001in
\subfigure[Prompt: Light painting photo of a cheetah, cinematic.]{\includegraphics[width=1\linewidth]{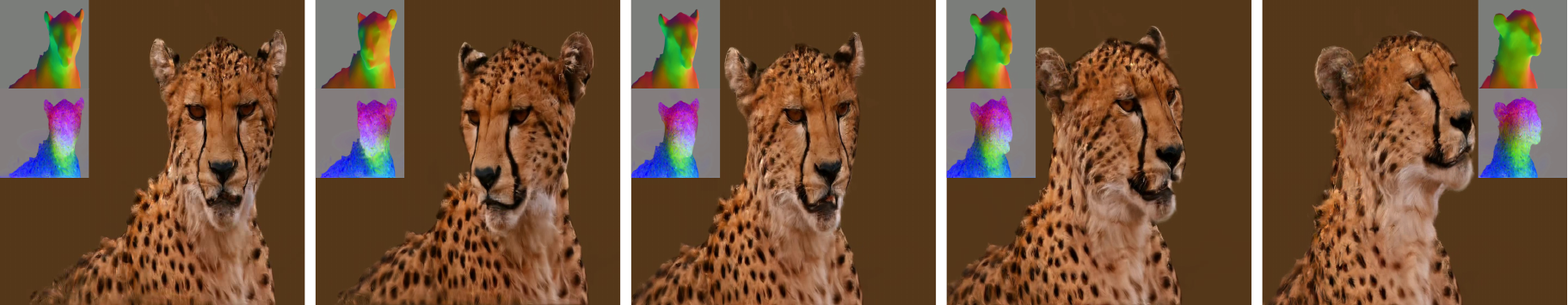}\label{fig:img_cheetah}} \vskip-0.001in
\subfigure[Prompt: A goldfish seemingly swimming through the air.]{\includegraphics[width=1\linewidth]{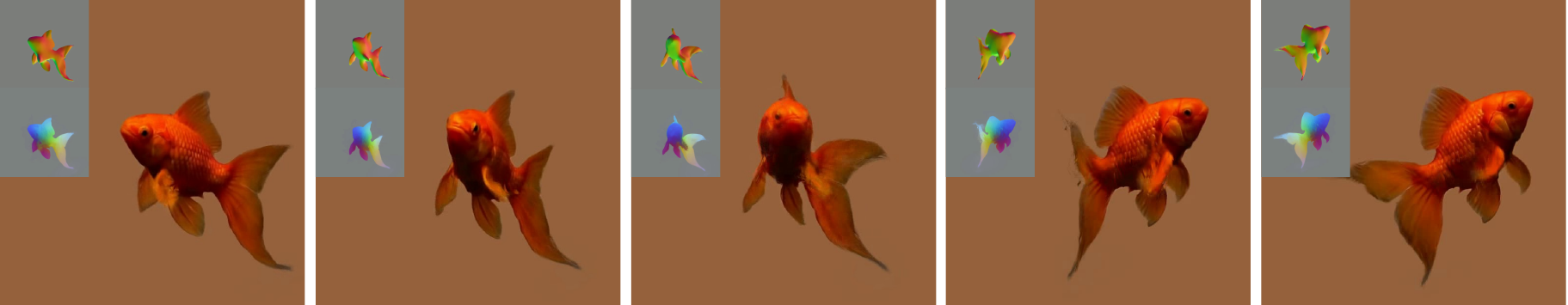}\label{fig:img_fish}} \vskip-0.001in
\subfigure[Prompt: A small, fluffy creature with an appearance reminiscent of a mythical being. The creature's fur texture is rendered in high detail. The monster's large eyes and open mouth express wonder and curiosity.]{\includegraphics[width=1\linewidth]{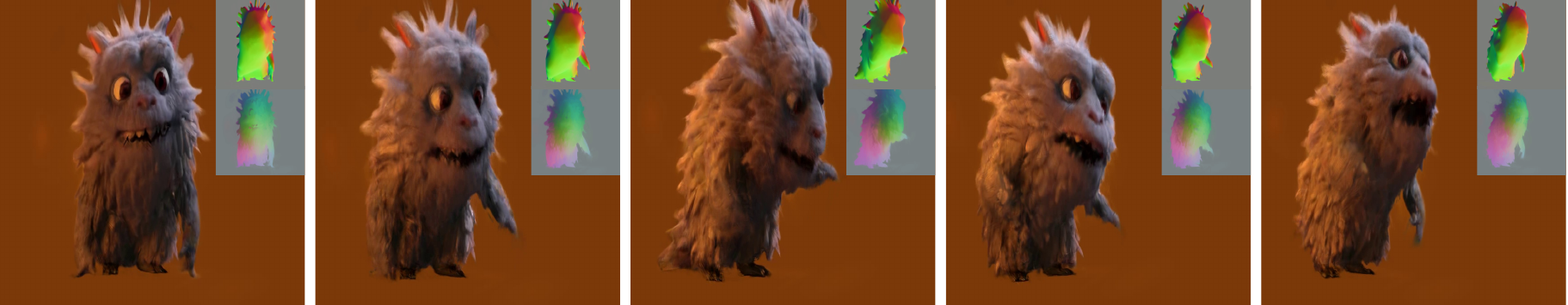}\label{fig:img_cheetah}} \vskip-0.001in
\subfigure[Prompt: An isolated coloured abstract sculpture with a dali shape.]{\includegraphics[width=1\linewidth]{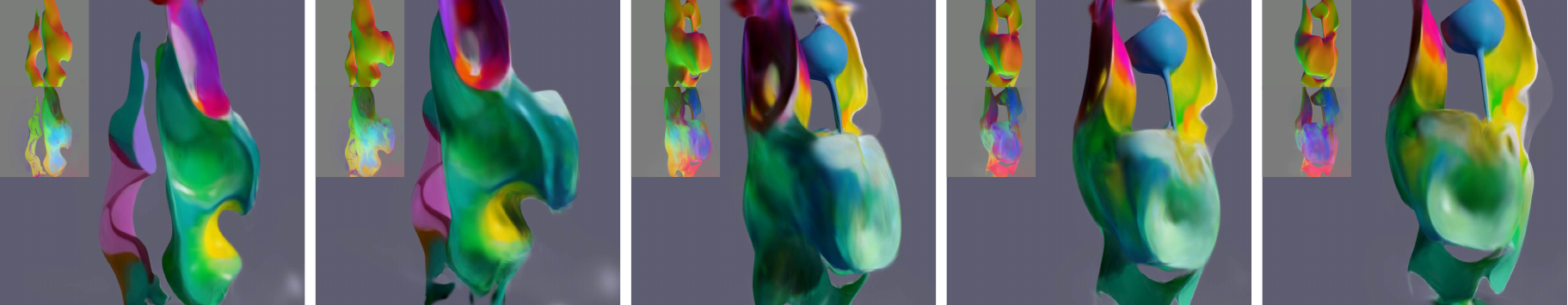}\label{fig:img_cheetah}}\vskip-0.001in
\subfigure[Prompt: An animated character  standing alongside a  dragon-like creature.]{\includegraphics[width=1\linewidth]{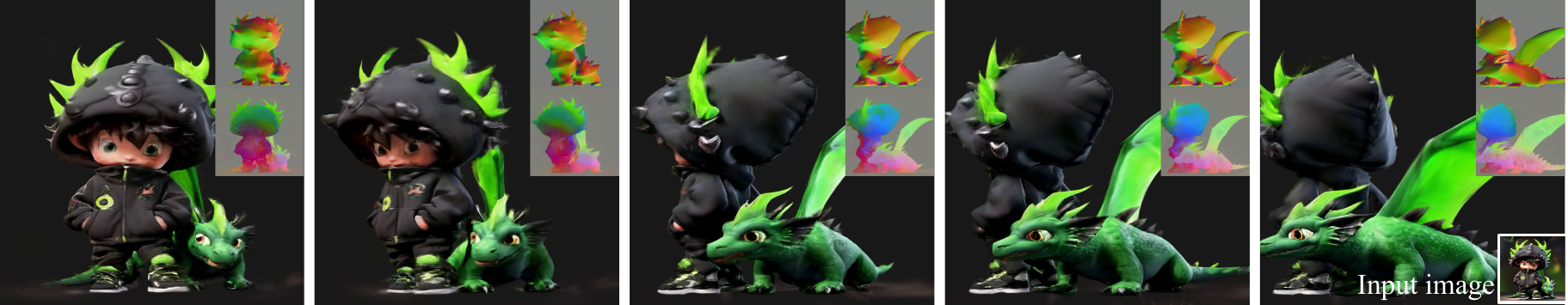}\label{fig:img_green_chara}}
\vskip-0.001in
\subfigure[Prompt: Vibrant, mythical phoenix  characterized by  bright orange feathers. ]{\includegraphics[width=1\linewidth]{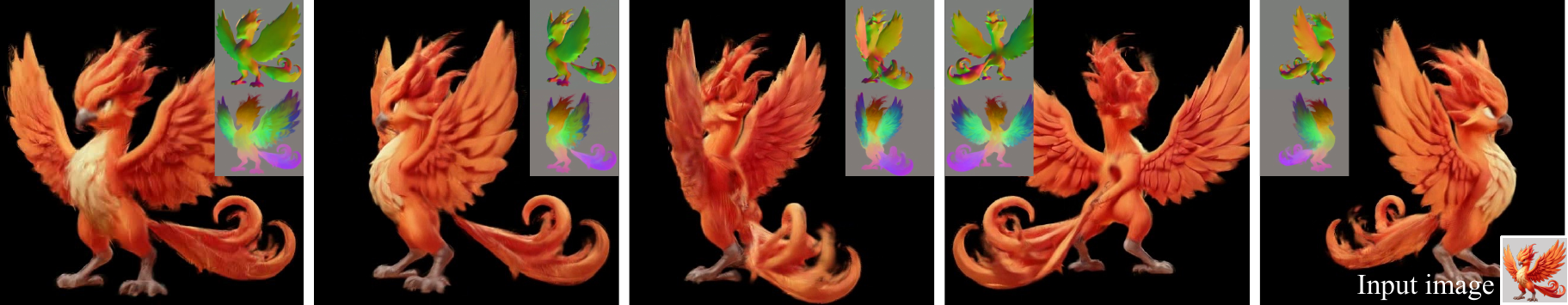}\label{fig:img_bird}}
\vskip-0.001in
\subfigure[Prompt: A cartoonish green dragon with orange-tinted horns and wings.]{\includegraphics[width=1\linewidth]{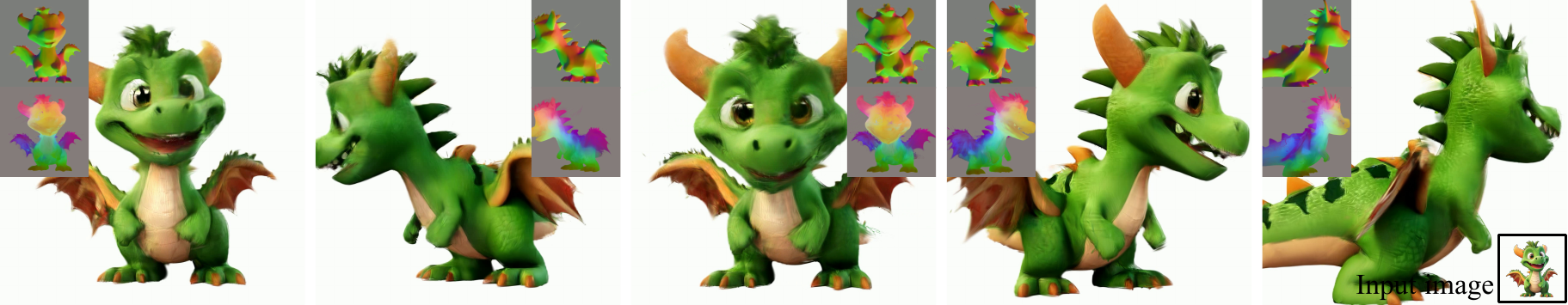}\label{fig:img_dinosaur}}
\vskip-0.001in
\caption{Video4DGen performance for \textbf{4D generation from generated videos}. For each sample, we present per-frame volume rendering for novel-view color,  normal, and surfel features. Video4DGen exhibits detailed and photo-realistic 4D representation. }\vskip-0.3in
\label{fig:intro-image}
\end{figure}

\begin{figure}[t]
\centering\vskip-0.046in\vskip-0.09in\hskip-0.01in
\subfigure[4D-guided ``multi-camera'' video generation. The three lines of each case follow the \textbf{same} motion sequence.
]{\includegraphics[width=1\linewidth]{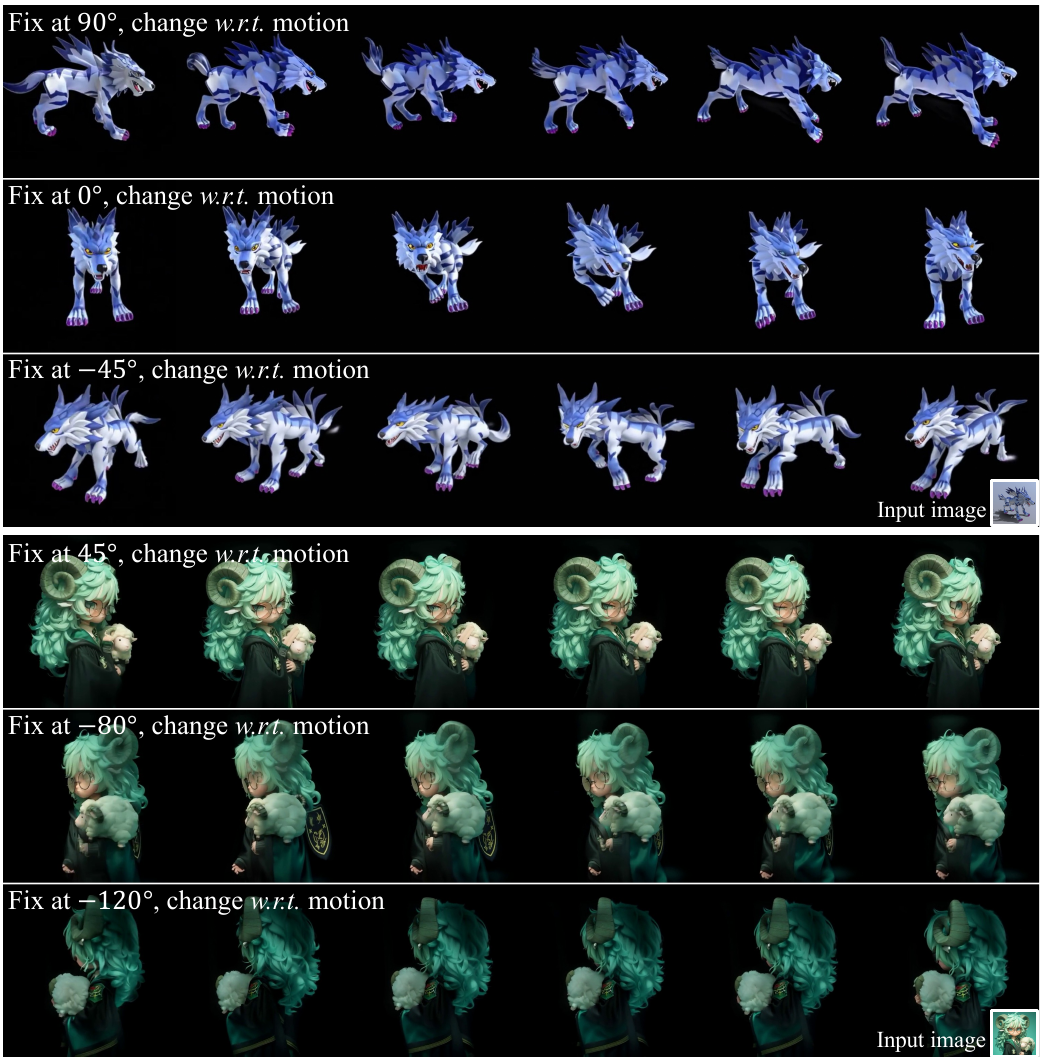}\label{fig:multicamera}}\vskip-0.01in
\subfigure[4D-guided video generation with large camera pose variations. The two lines of each case exhibit \textbf{different} motion sequences.
]{\includegraphics[width=1\linewidth]{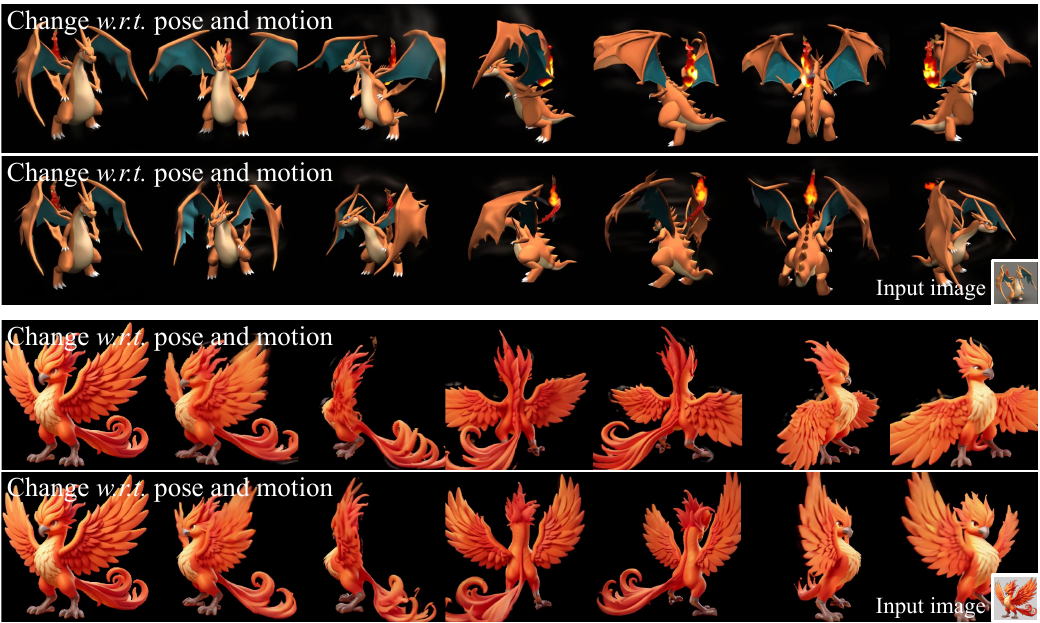}\label{fig:360}}
\caption{Video4DGen performance for \textbf{video generation with 4D guidance}. We present two novel video generation settings that are enabled by leveraging 4D generation, which is crucial for preserving coherent motion sequences across both viewpoints and time. }\vskip-0.18in
\label{fig:intro-image-video}
\end{figure}

{Despite these advancements, achieving high-fidelity 4D reconstruction based on video generative models poses great challenges. One of the primary issues lies in the non-rigidity and frame distortion usually found in generated videos, which  can undermine the temporal consistency and spatial coherence of the generated 4D content. Applying existing 4D methods~\cite{pumarola2021d,yang2023deformable3dgs,huang2023sc} to generated videos usually struggle to maintain smooth transitions across frames and viewpoints, leading to distortion and artifacts such as flickering and misalignment. Furthermore, generating plausible 4D content from novel viewpoints, particularly for regions not captured in the original input, \emph{i.e.}, the generated video, remains an unresolved challenge. Additionally, most generated videos lack the explicit camera pose information, whereas existing state-of-the-art 4D  methods~\cite{DBLP:journals/tog/ParkSHBBGMS21,wu20234dgaussians,yang2023deformable3dgs,huang2023sc,DBLP:conf/cvpr/LiCLX24} require accurate camera poses to align the spatial-temporal structure properly.}

In response to these challenges,  we present \textbf{Video4DGen}, a multimodal dynamic generation framework for jointly performing 4D generation  and 4D-guided video generation. Video4DGen optimizes appearance and geometry across a wide range of pose and motion dimensions, ensuring spatial and temporal  consistency throughout the generation process. The  framework introduces  a novel 4D representation \textbf{Dynamic Gaussian Surfels (DGS)}, enhanced by a specially designed field initialization. It features two key stages: 4D generation from single or multiple generated videos and novel-view video generation guided by the 4D representation.

Specifically, the proposed DGS optimizes non-rigid warping functions that transform Gaussian surfels from static to dynamically warped states. This dynamic transformation accurately represents motion and deformation over time, crucial for capturing realistic 4D geometry and appearance. Besides, DGS demonstrates superior 4D representation performance due to two other key aspects. Firstly, in terms of geometry, DGS adheres to Gaussian surfels principles~\cite{Huang2DGS2024,Dai2024GaussianSurfels} to achieve precise geometric representation. Unlike existing methods, DGS incorporates warped-state normal consistency regularization to align surfels with actual surfaces with learnable continuous fields (\emph{w.r.t.} spatial coordinate and time) to ensure smooth warping when estimating normals. Secondly, for appearance, DGS learns additional refinements on the rotation and scaling parameters of Gaussian surfels by a dual branch structure. This refinement reduces the flickering artifacts during warping and allows for the precise rendering of appearance details, resulting in high-quality  4D representations. 

Since camera trajectories for generated videos are unknown, Structure from Motion (SfM) techniques like COLMAP~\cite{schonberger2016structure,DBLP:conf/eccv/SchonbergerZFP16} often encounter difficulties in converging due to the presence of non-rigid deformations. To overcome this challenge, we introduce  field initialization as a critical component in our pipeline. Field initialization is designed to set up the continuous warping field of DGS, ensuring rapid and stable convergence. With this initialization, Video4DGen is able to achieve high-fidelity 4D generation from single generated videos.

When extending 4D generation from single to multiple generated videos, we design several key mechanisms including static-state sharing, continuous root pose optimization, and pose-guided frame sampling. Static-state sharing maintains consistency in the subject's structure and appearance by preserving a consistent set of DGS for the static state. Root pose optimization aligns the global transformations of the subject across frames with the static state, correcting mismatches and ensuring smooth transitions. Video4DGen utilizes neural continuous fields for both local and global pose adjustments. In addition, pose-guided frame sampling enhances pose diversity while avoiding overfitting  certain poses during training. The multi-video generation  could  improve the quality of 4D generation across the full spectrum of  views.

Video4DGen also introduces a novel 4D-guided video generation approach that enhances the video generation process by integrating diverse combinations of motion and viewpoints. It utilizes two key strategies. Firstly, a confidence-filtered DGS mechanism uses normal alignment to assess pixel reliability and only includes high-confidence regions in the output. Secondly, novel-view video generation refines these high-confidence regions with progressive denoising, while low-confidence areas undergo transformation to ensure smooth transitions. Together, these strategies improve the overall clarity and coherence of the generated videos. Empowered by the generated 4D, our video generator exhibits novel features such as ``multi-camera'' video generation where the same motion sequence is simultaneously captured from any viewpoints and  video generation with large pose changes. The 4D-guided video generation in turn enhances the 4D generation.

We provide  visualization results for 4D generation in  Fig.~\ref{fig:intro-image} and 4D-guided video generation in Fig.~\ref{fig:intro-image-video}.

Extensive experiments based on the generated videos verify the effectiveness of our method compared to existing state-of-the-art methods.  We also provide quantitative and qualitative comparisons on object-level benchmarks and realistic scene-level benchmarks. Therefore, our framework is not limited to only representing  objects  or object-centric scenes. Results demonstrate the superior performance of our framework in terms of both visual quality and geometry details.

In summary, we propose Video4DGen, a novel framework for multimodal dynamic generation, capable of performing both 4D generation and 4D-guided video generation. The main contributions of Video4DGen include:
\begin{itemize} 

\item \textbf{DGS as high-fidelity 4D representation.} As the key 4D representation of Video4DGen, the designed  DGS   captures non-rigid motion and deformation with continuous fields,   enhancing both geometric precision and visual quality.

\item \textbf{4D generation from generated videos.} Video4DGen performs 4D generation from single or multiple generated videos, with  a  field initialization strategy to facilitate the stable convergence of DGS. We develop various techniques including static-state sharing,  root pose optimization and frame sampling to further enhance spatial and temporal consistency.

\item \textbf{Video generation with 4D  guidance.}  Using filtered DGS as  4D guidance, Video4DGen  excels in producing novel-view video generation with controllable camera transitions. And as far as we know, it is the first framework to achieve  multi-camera video generation with 4D guidance.
\end{itemize}

\section{Related works}

{This section examines three research domains fundamental to our work, including 3D representations, 4D reconstruction/generation, and 3D/4D-guided video generation. We systematically analyze technological trajectories and identify critical gaps that our framework addresses.}

\textbf{3D representation.} Transforming 2D images into 3D representations has long been a central challenge in the field. Initially, triangle meshes were favored for their compactness and compatibility with rendering pipelines \cite{buehler2001unstructured, debevec1996modeling, waechter2014let, wood2000surface, riegler2020free, thies2019deferred}. However, the transition to more sophisticated volumetric methods was inevitable due to the limitations of surface-based approaches. Early volumetric representations included voxel grids \cite{sitzmann2019deepvoxels, lombardi2019neural, penner2017soft, kutulakos2000theory} and multi-plane images \cite{zhou2018stereo, flynn2019deepview, mildenhall2019local, srinivasan2019pushing, srinivasan2020lighthouse, tucker2020single}, which, despite their straightforwardness, demanded intricate optimization strategies. The introduction of neural radiance fields (NeRF) \cite{DBLP:conf/eccv/MildenhallSTBRN20} marked a significant advancement, offering an implicit volumetric neural representation that could store and query the density and color of each point, leading to highly realistic reconstructions. The NeRF paradigm has since been improved upon in terms of reconstruction quality \cite{barron2021mip, barron2023zipnerf, kerbl3Dgaussians, ma2021deblur, wang20224k} and rendering~\cite{reiser2023merf, hedman2021snerg, yu2021plenoctrees, reiser2021kilonerf, liu2020neural, rebain2021derf, lindell2021autoint, garbin2021fastnerf, Hu_2022_CVPR, cao2022mobiler2l, wang2022r2l, lombardi2021mixture}.  To address the limitations of NeRF, such as rendering speed and memory usage, recent work dubbed 3D Gaussian splatting~\cite{kerbl3Dgaussians} has proposed anisotropic Gaussian representations with GPU-optimized tile-based rasterization. This has opened up new avenues for surface extraction \cite{Huang2DGS2024, guedon2023sugar}, generation \cite{chen2023text, tang2024lgm, xu2024grm}, and large-scale scene reconstruction \cite{liu2024citygaussian, shuai2024LoG, hierarchicalgaussians24}. Gaussian surfels methods~\cite{Huang2DGS2024,Dai2024GaussianSurfels} further exhibit advantages in modeling accurate geometry. While these methods have  advanced the field of static 3D representation, capturing  dynamic aspects of real-world scenes/subjects with non-rigid motion and deformation introduces a distinct set of challenges that demand new solutions.

\textbf{4D reconstruction/generation.}
Extending static 3D reconstruction to spatiotemporal domains introduces complex challenges in motion decomposition and temporal consistency, necessitating the capture of non-rigid motion and deformation over time \cite{li2022tava, peng2021neural, su2021anerf, zhang2021stnerf, wang2023animatabledreamer}. Traditional methods have explored dynamic reconstruction using synchronized multi-view videos \cite{lombardi2019neural, DBLP:conf/nips/0011SW0T22, wang2023rpd, attal2023hyperreel, song2022nerfplayer, cao2023hexplane, wang2022mixed, wang2022fourier, bansal20204d, peng2023representing, wang2023masked} or have focused on specific dynamic elements like humans or animals. More recently, there has been a shift towards reconstructing non-rigid objects from monocular videos, which is a more practical yet challenging scenario. One approach involves incorporating time as an additional input to the neural radiance field \cite{DBLP:conf/cvpr/LiSZGL0SLGNL22, kplanes_2023, cao2023hexplane,DBLP:conf/cvpr/YangVNRVJ22}, allowing for explicit querying of spatiotemporal information. Another line of research decomposes the spatiotemporal radiance field into a canonical space and a deformation field, representing spatial attributes and their temporal variations \cite{pumarola2021d, TiNeuVox, DBLP:journals/tog/ParkSHBBGMS21, dycheck, DBLP:conf/iccv/ParkSBBGSM21, TiNeuVox, shao2023tensor4d, liu2023robust, liu2022devrf, Zhao_2022_CVPR, tretschk2021non, jiang2022alignerf, du2021nerflow, gao2021dynamic, li2020neural, xian2021space}. With advancements in Gaussian splatting, deformable-GS \cite{yang2023deformable3dgs} and 4DGS \cite{wu20234dgaussians} have been developed, utilizing neural deformation fields with multi-layer perception (MLP) and triplane, respectively. SC-GS~\cite{huang2023sc} and dynamic 3D Gaussians \cite{luiten2023dynamic} also advance the field by modeling time-varying scenes. {4D-Rotor~\cite{DBLP:conf/siggraph/DuanWDHCC24} introduces anisotropic 4D Gaussians with rotor-based rotation representations and temporal slicing to model complex dynamic scenes.} In the realm of 3D or 4D generation, our 4D generation pipeline diverges from recent progress in optimization-based \cite{poole2022dreamfusion, wang2023prolificdreamer, lin2023magic3d, chen2023fantasia3d, chen2023text,wang2023animatabledreamer,singer2023text,ling2023align,bah20244dfy}, feed-forward \cite{hong2023lrm, zou2023triplane, wang2024crm}, and multi-view reconstruction methods \cite{chen2024v3d, long2023wonder3d, lu2023direct2} by leveraging a video generative model to achieve generation capabilities. In this paper, one primary focus is preserving high-quality appearance and geometrical integrity from generated videos. {Despite advancements in existing approaches~\cite{DBLP:conf/iccv/ParkSBBGSM21,pumarola2021d,DBLP:conf/cvpr/YangVNRVJ22,wu20234dgaussians,yang2023deformable3dgs,huang2023sc,DBLP:conf/cvpr/LiCLX24,DBLP:conf/siggraph/DuanWDHCC24}, applying   them to reconstructing 4D contents from generated videos exhibits three critical limitations: susceptibility to appearance/geometric inconsistencies from video generation artifacts, dependence on known camera poses, and limited multi-video alignment capabilities. Our DGS design stands out in its ability to perform 4D reconstruction from generated videos by several key innovations. 
Firstly, DGS incorporates warped-state regularization and a dual-branch structure, effectively addressing the geometry and appearance inconsistencies commonly found in the generated videos. Secondly, with our field initialization, DGS  handles large object deformations and root pose changes even in the absence of known camera poses. This capability is crucial since generated videos often lack explicit camera pose information. Thirdly, DGS maintains a shared static state while learning non-rigid warping for Gaussian surfels, enabling robust alignment across multiple generated videos and enhancing the quality of multi-video 4D reconstruction. These together result in a generation process that not only captures the  motion and deformation,  but also maintains  high standards of geometric and appearance  details, essential for creating immersive and lifelike virtual  representations. }

\textbf{3D/4D-aware video generation.} The integration of video generation with  3D or 4D guidance has given rise to a novel class of methods that leverage  structural information to enhance video content. For instance, VD3D~\cite{DBLP:journals/corr/abs-2407-12781} tames large-scale video diffusion transformers to allow precise 3D camera control during video generation, an essential advancement for applications requiring accurate spatial manipulation. CamCo~\cite{DBLP:journals/corr/abs-2406-02509} builds on this idea by generating 3D-consistent videos from images, enhancing video content with dynamic camera movements. Similarly, CVD~\cite{DBLP:journals/corr/abs-2405-17414} generates multi-video sequences and maintains consistency across camera controls, a key for multi-view dynamic video generation. Existing methods have also utilized posed datasets to guide the generation of videos from novel viewpoints~\cite{DBLP:journals/corr/abs-2404-02101}. This approach facilitates the creation of smooth transitions and ensures consistency in the appearance of objects across frames. Furthermore, for existing methods, while employing 4D models to direct video generation captures the temporal dynamics of the content~\cite{DBLP:conf/cvpr/CaiCGHWW24}, it falls short in offering detailed guidance or accommodating 4D changes with varying poses. Unlike existing methods, we present  Video4DGen as a novel framework that jointly optimizes video and 4D generation. {One of the key innovations  lies in the specialized design of DGS. Apart from its specialty of 4D generation from generated videos as mentioned above,  DGS introduces a novel application by enhancing Gaussian surfels as confidence-filtered guidance which effectively improves the visual quality of 4D-guided video generation. This also differs from existing 4D approaches, marking a step forward in dynamic content creation. }

\section{Preliminary}
\label{subsec:preliminary} 

\subsection{Video Diffusion Model} 
\label{subsec:preliminary_video} 
We consider a basic video diffusion model to model the transformation of video frames over time. Suppose the video diffusion model contains an encoder $\text{Enc}(\cdot)$ and a decoder $\text{Dec}(\cdot)$. During training, $\text{Enc}(\cdot)$ learns to encode each video data, denoted by $\Vv \in \mathbb{R}^{F \times H \times W \times 3}$, with $F$ being the number of video frames and $H\times W$ indicating the  resolution per frame. The video latent $\z_0 = [\gz^1_0;\text{\myldots};\gz^F_0]=\text{Enc}(\Vv)\in \mathbb{R}^{F \times H' \times W' \times C}$ with  $H' \times W'$ indicating the compressed  spatial dimensions and $C$ denoting the number of the latent channel. Then $\text{Dec}(\cdot)$ decodes from the latent $\z_0$ to obtain the generated video as $\text{Dec}(\z_0) \in \mathbb{R}^{F \times   H \times   W \times 3}$. To facilitate understanding, we maintain the temporal length of the latent representation as $F$. Our framework is also compatible with video diffusion models that incorporate temporal compression.  
For  a diffusion  time step schedule $0=\tau_0 < \tau_1< \text{\mycdots} <\tau_{S}=T$ initialized by a diffusion scheduler, the model generates a video by iteratively denoising from the start $\z_{\tau_S}=[\gz_{\tau_S}^1;\text{\myldots};\gz_{\tau_S}^F] \sim \mathcal{N}(\mathbf{0}, \mathbf{I})$  for $S$ times using a sampler $\Phi(\cdot)$, \emph{e.g.}, the DDIM sampler~\cite{DBLP:conf/nips/SongE19}. Each denoising step is performed by
\begin{align}\label{eq:diffusion_step}
   \z_{\tau_{t-1}}= [\gz_{\tau_{t-1}}^1;\text{\myldots};\gz_{\tau_{t-1}}^F] = \Phi\big([\gz_{\tau_{t}}^1;\text{\myldots};\gz_{\tau_{t}}^F], \bm{c}; \bv_\Theta\big),
\end{align}
where $\gz_{\tau_t}^f$ ($f=1,\text{\myldots},F$) is the latent embedding at time step $\tau_t$ for the $f^\text{th}$ frame (or the $f^\text{th}$ temporal dimension of the latent), $\bm{c}$ is the text/image condition, and $\bv_\Theta(\z_{\tau_{t}},\tau_{t})$ refers to the velocity predicted by a neural network parameterized by  weights $\Theta$.  In the following, we omit the condition $\bm{c}$  for simplicity. 

\subsection{4D Reconstruction} 
Given an input video with $F$ frames, the goal of 4D reconstruction is to determine a sequential 3D representation that could be rendered to fit each video frame as much as possible. Specifically, suppose the 3D representation for the $f^\text{th}$ frame  is parameterized by $\theta_f$, where $f=1,\cdots,F$.  Given a differentiable rendering mapping  $\vg$, we could obtain the rendered color at the frame pixel $\bar\bx^f\in\mathbb{R}^2$. We choose volume rendering as commonly adopted in NeRF~\cite{DBLP:conf/eccv/MildenhallSTBRN20}, Gaussian splatting~\cite{kerbl3Dgaussians}, and Gaussian surfels~\cite{Huang2DGS2024,Dai2024GaussianSurfels}. The optimization of 4D reconstruction can be implemented by minimizing the empirical loss as  
\begin{equation}
\label{eq:loss}
    \min_{\theta}\frac{1}{F}\sum_{f=1}^{F}\sum_{\bar\bx^f}\mathcal{L}\Big(\br(\bar\bx^f)=\vg\big(\theta_f, \{\bx^f_i\}_{i=1,\cdots,I}\big),  \hat\br(\bar\bx^f)\Big),
\end{equation}
where  $\mathcal{L}$ refers to a supervision loss, \emph{e.g.}, L2 loss; $\bx^{f}_i \in \mathbb{R}^3$ is the $i^\text{th}$ 3D point sampled or intersected with Gaussian primitives along the ray that emanates from the frame pixel $\bar\bx^{f}$; $I$ is the number of sampled or intersected points per ray; $\br(\bar\bx^{f})$ and $\hat\br(\bar\bx^{f})$ are the rendered color and the observed color at  $\bar\bx^{f}$, respectively.

\section{Video4DGen}
In this work, we present a framework Video4DGen which contains a novel 4D representation, 4D generation from generated videos, and 4D-guided video generation. We start by outlining the foundational problem definition  in Sec.~\ref{subsec:definition}. Subsequently, we design  DGS  in Sec.~\ref{subsec:modeling} to precisely represent both the visual and geometrical characteristics of generated videos. In Sec.~\ref{subsec:Video4DGen}, we design the field initialization process, which creates and initializes an implicit field to refine  poses and the warping field for motions. We propose 4D generation from multiple generated videos for a wider range of camera perspectives in Sec.~\ref{sec:4dgen_multi}, and the corresponding video generation based on  4D guidance in Sec.~\ref{sec:vidgen}.

\subsection{Problem Definition}
\label{subsec:definition} 

We now provide the overall formulation for each of the  primary components in Video4DGen, \emph{i.e.}, the 4D representation DGS, 4D generation, and 4D-guided video generation. Suppose for each case, DGS is built based on a total of $M$ generated videos ($M$ could be 1) with each video consisting of $F$ frames. 

\textbf{DGS as high-fidelity 4D representation.} We maintain parameters in both static state and per-frame warped state, by
\begin{equation}
\bm{\theta}^{*}=\{\bo_k^*,\bR_k^*,\bS_k^*,\bc_k,\alpha_k\}_{k=1}^K,\quad \bJ=\{\bJ^{f,m}\}_{f=1,m=1}^{F,M},
\label{eq:overall_dgs}
\end{equation}
where $\bm{\theta}^{*}$ represents  the 4D parameters in the static state and $\bJ$ refers to the set of local rigid transformations over $\bm{\theta}^{*}$, with $\bJ^{f,m}=[
\tilde\bR^{f,m},\tilde\bT^{f,m}]\in\SE(3)$. Details including parameter definitions will be  introduced in Sec.~\ref{subsec:modeling}.

\textbf{4D generation from generated videos.} Similar to Eq.~(\ref{eq:loss}), we optimize $\bm{\theta}^{*}$ and $\bJ$ here to  minimize the loss written as
\begin{equation}
\min_{\bm{\theta}^{*},\bJ}\mathbb{E}_{\{f,m\}\sim\phi}\sum_{\bar\bx^{f,m}}\mathcal{L}\Big(\br(\bar\bx^{f,m}), \text{Dec}(\z_0)(\bar\bx^{f,m})\Big),
\label{eq:overall_4dgen}
\end{equation}
where $\br(\bar\bx^{f,m})$ is the rasterized result of DGS along the camera ray that emanates from the frame pixel $\bar\bx^{f,m}$; $\text{Dec}(\z_0)(\bar\bx^{f,m})$  is the  observed color  of the generated video at  $\bar\bx^{f,m}$.  $\phi$ refers to a pose-guided frame and video sampling strategy. We will describe this part in Sec.~\ref{sec:4dgen_multi}.

\textbf{Video generation with 4D  guidance.} Once we obtain a 4D representation with a wide range of camera perspectives, we could perform 4D-guided video generation. Each denoising step of 4D-guided video generation is performed by
\begin{align}\label{eq:diffusion_step}
\z_{\tau_{t-1}}^m&= [\gz_{\tau_{t-1}}^{1,m};\text{\myldots};\gz_{\tau_{t-1}}^{F,m}]\\&= \Phi'\big([\gz_{\tau_{t}}^{1,m};\text{\myldots};\gz_{\tau_{t}}^{F,m}]; \bm{\theta}^{*}; [\bJ^{1,m};\text{\myldots};\bJ^{F,m}];\bv_\Theta\big),
\label{eq:overall_videogen}
\end{align}
where $\Phi'$ refers to a mixed denoising process with several specific forms according to the flow trajectory of  the video diffusion model. The denoising process also involves filtering the generated DGS with a confidence map. We provide more details in Sec.~\ref{sec:vidgen}. 

\begin{figure*}[t]
    \centering\vskip0.03in
     \hskip-0.05in 
\includegraphics[width=1.8\columnwidth]{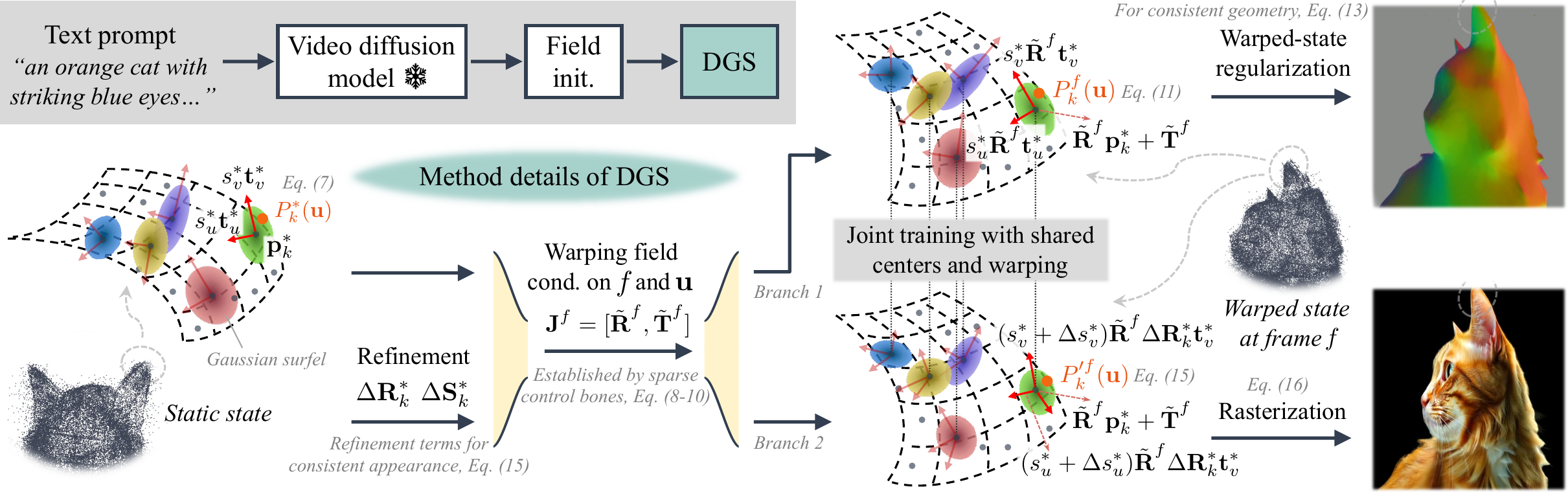}
    \vskip-0.01in
\caption{{Illustration of our 4D generation framework with  DGS in detail.}  {For DGS, Gaussian surfels in the static state are transformed to the warped state by learning the non-rigid warping field which is conditioned on the frame $f$ and coordinate $\bu$. The local warping at each Gaussian surfel is a $\SE(3)$ transformation, established by sparse control bones.  DGS incorporates warped-state geometric regularization and a dual-branch refinement strategy, which address geometry and appearance  inconsistencies, respectively. Both branches share the same centers of Gaussian primitives and the same warping field. The second branch further refines the rotations and scaling matrices of Gaussian primitives  to enhance detailed appearance.} ``Field init.'' stands for field initialization as introduced in Sec.~\ref{subsec:Video4DGen}.}
\label{fig:method}
\end{figure*}

\subsection{Dynamic Gaussian Surfels}
\label{subsec:modeling} 

In this part, we first consider the 4D representation and generation from a single generated video, namely, $M=1$. Hence,  the video index $m$ is omitted here. We start to perform 4D generation from multiple generated videos later in Sec.~\ref{sec:4dgen_multi}.

Essentially, our goal is to build a sequential  3D representation that could deform to be consistent with each 2D frame. We begin at considering an ideal video exhibiting different views of the same static object without object deformation, movement, or video distortion. To model the 3D  representation with high appearance fidelity and geometry accuracy, we follow the method of using differentiable 2D Gaussian primitives as proposed by recent Gaussian surfel advances~\cite{Huang2DGS2024,Dai2024GaussianSurfels}. Specifically, the  $k^\text{th}$ Gaussian surfel (of the total $K$)  is characterized by a central point $\bp_k^*\in\mathbb{R}^3$, a local coordinate system centered at $\bp_k^*$ with  two principal tangential vectors $\bt_u^*\in\mathbb{R}^{3\times 1}$, $\bt_v^*\in\mathbb{R}^{3\times 1}$, and scaling factors $s_u^*\in\mathbb{R}$, $s_v^*\in\mathbb{R}$. Here, we use the notation ``$*$'' to represent parameters in the static state. A Gaussian surfel is computed as a 2D Gaussian defined in a local tangent plane in the world space. Following~\cite{Huang2DGS2024}, for any point $\bu=(u,v)$ located on the  $uv$ coordinate system centered at $\bp_k^*$, its coordinate  in the world space, denoted as $P_k^*(\bu)\in\mathbb{R}^{3\times 1}$, is computed by
\begin{align}
P_k^*(\bu)=\bp_k^*+s_u^*\bt_u^*u+s_v^*\bt_v^*v=\begin{bmatrix}
        \bR_k^*\bS_k^* & \bp_k^* 
    \end{bmatrix}(u,v,1,1)^\top,
\label{eq:2dgaussian}
\end{align}
where  $\bR_k^*=[\bt_u^*,\bt_v^*, \bt_u^*\times \bt_v^*]\in\SO(3)$ denotes the rotation matrix, and the diagonal matrix $\bS_k^*=\mathrm{diag}(s_u^*, s_v^*, 0)\in\mathbb{R}^{3\times 3}$ denotes the scaling  matrix.

Instead of modeling static 3D objects, in this work, we aim to  develop a presentation strategy for generating 4D content from generated videos that may  exhibit large non-rigidity, distortion, or illumination changes. We introduce \textbf{Dynamic Gaussian Surfels (DGS)}, a representation designed to achieve precise 4D generation while accommodating non-rigidity and other time-varying effects.

Motivated by recent advancements in non-rigid reconstruction methods~\cite{DBLP:conf/iccv/ParkSBBGSM21,DBLP:conf/cvpr/YangVNRVJ22,wang2023animatabledreamer},  we aim to ensure that the target object maintains a consistent static state across different frames, thereby mitigating non-rigidity and distortion effects. To achieve this, we employ warping techniques on each Gaussian surfel represented by  $P_k^*(\bu)$, transforming it into a corresponding Gaussian surfel $P_k^f(\bu)$ at the $f^\text{th}$  frame, which is centered at $\bp^f_k\in\mathbb{R}^3$ with a rotation matrix $\bR_k^f\in\SO(3)$ and a scaling matrix $\bS_k^f\in\mathbb{R}^{3\times 3}$.

\textbf{Non-rigid warping for Gaussian surfels.} 
We now build the warping process from the static state to the warped state. We define a time-varying non-rigid warping function by leveraging $B$  bones as key points to ease the training of deformation.  In the static state, the $b^\text{th}$ bone is represented by 3D Gaussian ellipsoids~\cite{DBLP:conf/cvpr/YangSJVCCRFL21} with the center ${\bc}^*_b\in\mathbb{R}^{3 \times 1}$, rotation matrix $\bV_b^*\in\mathbb{R}^{3\times3}$, and  diagonal scaling matrix $\boldsymbol{\Lambda}_b^*\in\mathbb{R}^{3\times3}$.  We let $\bJ_b^{f}\in\SE(3)$ represent a rigid transformation that moves the $b^\text{th}$ bone from its static state to the warped state at the $f^\text{th}$  frame.  For a 3D point $P_k^*(\bu)$, the skinning weight vectors $\bw^{f}\in\mathbb{R}^{B\times1}$  at the $f^\text{th}$  frame is calculated by the normalized Mahalanobis distance following~\cite{DBLP:conf/cvpr/YangVNRVJ22}
\begin{align}
    {\delta}^{f}_b &= \big(P_k^*(\bu)-{\bc}^{f}_b\big)^\top{\bf Q}^{f}_b\big(P_k^*(\bu)-{\bc}^{f}_b\big),\\\quad\bw^{f}&=\sigma_\mathrm{softmax}\big({\delta}^{f}_1, {\delta}^{f}_2, \cdots,{\delta}^{f}_B\big)^\top,
 \label{eq:bone}
\end{align}
where ${\delta}^{f}_b$ computes the squared distance between $P_k^*(\bu)$ and the $b^\text{th}$ bone; ${\bc}^f_b\in\mathbb{R}^{3 \times 1}$ is the center of the $b^\text{th}$ bone at the $f^\text{th}$  frame,  and ${\bf Q}^{f}_b = {\bV_b^{f}}^\top\boldsymbol{\Lambda}_b^*\bV_b^{f}$ is the precision matrix composed by the bone orientation matrix $\bV_b^f\in\mathbb{R}^{3\times3}$ and $\boldsymbol{\Lambda}_b^*$. Specifically, there is $(\bV_b^{f}|{\bc}^{f})=\bJ_b^{f}(\bV_b^*|{\bc}^*)$ with ${\bc}^*_b$, $\bV_b^*$, and $\boldsymbol{\Lambda}_b^*$ being learnable parameters. $\sigma_\mathrm{softmax}$ is the $\mathrm{softmax}$ function.

In effect, $\mathbf{J}^{f}_{b}$ is achieved by non-linear mappings using a multi-layer perception (MLP) with $\SE(3)$ guaranteed, as will be given  in Eq.~(\ref{eq:nerf_bone}). The non-rigid warping function is a weighted combination of $\bJ_b^{f}\in\SE(3)$, where we apply dual quaternion blend skinning (DQB)~\cite{DBLP:conf/si3d/KavanCZO07} to ensure valid $\SE(3)$ after combination,
\begin{align}
\bJ^{f}=\mathcal{R}\Big(\sum_{b=1}^{B} w_{b}^{f} \mathcal{Q}(\mathbf{J}^{f}_{b})\Big),
\label{eq:warpingJ1}
\end{align}
where $w_{b}^{f}$ is the $b^\text{th}$ element of $\bw^{f}$ calculated in Eq.~(\ref{eq:bone}); $\mathcal{Q}$ and $\mathcal{R}$ denote the quaternion process and the inverse quaternion process, respectively. In this case, $\bJ^{f}\in\SE(3)$.

We therefore rewrite the warping  as $\bJ^{f}=[
\tilde\bR^{f},\tilde\bT^{f}] $ with the rotation $\tilde\bR^{f}\in\SO(3)$ and translation $\tilde\bT^{f}\in\mathbb{R}^{3}$, and apply the corresponding transformation to Eq.~(\ref{eq:2dgaussian}) by 
\begin{align}
P_k^{f}(\bu)=\bJ^{f}P_k^*(\bu)=\begin{bmatrix}
        \tilde\bR^{f}\bR_k^*\bS_k^* & \tilde\bR^{f}\bp_k^*+\tilde\bT^{f} \\
    \end{bmatrix}(u,v,1,1)^\top.
    \label{eq:warping}
\end{align}

Note that Eq.~(\ref{eq:warping}) holds for any given point $P_k^*(\bu)$ including the  center point of the $k^\text{th}$ Gaussian surfel (\emph{i.e.}, $\bp_k^*$) when $\bu=(0,0)$. By deriving  Eq.~(\ref{eq:warping}), we enable connection of the warping function \emph{w.r.t.} to any point $\bu=(u,v)$ on the local coordinate system centered at $\bp_k^*$, which is needed later in Eq.~(\ref{eq:2dgs}) where $\bu$ is an intersection with Gaussian surfels and a ray that emanates from the frame pixel.

\textbf{Addressing geometry inconsistency.} 
To accurately capture the geometric representation, we follow similar methods in Gaussian Surfels~\cite{Huang2DGS2024,Dai2024GaussianSurfels} to add normal consistency regularization which encourages all Gaussian surfels to be locally aligned with the actual surfaces. Differently, unlike 3D representation for static scenes, 4D representation commonly faces non-rigidity and distortion. Thus simply performing regularization to promote surface-aligned Gaussian surfels like previous methods harms the structural integrity due to the non-rigid warping.

We therefore design a warped-state geometric regularization. As mentioned in Eq.~(\ref{eq:warping}),  each point $P_k^{f}(\bu)$ in the warped state (frame  $f$) is transformed from its corresponding static point $P_k^*(\bu)$ based on the warping function, \emph{i.e.}, $P_k^{f}(\bu)=\bJ^{f}P_k^*(\bu)$ with $\bJ^{f}$ composed by  $\bJ_b^{f}$. To maintain the structural integrity to a large extent when addressing geometry inconsistency, we design $\bJ_b^{f}$ as a continuous field that takes both the point $P_k^*(\bu)$ (or equivalently, $\bu$ in the local coordinate system) and the frame $f$ as conditions. By this setting, $\bJ_b^{f}$ is expected to change continuously with the change of $\bu$ or $f$. We implement the continuous field by using a NeRF-style MLP which directly outputs a 6-dimensional dual quaternion, and rely on  the inverse quaternion process $\mathcal{R}$ to guarantee $\SE(3)$, \emph{i.e.}, 
\begin{equation}
{\bf J}^{f}_b = \mathcal{R}\big(\mathbf{MLP}(\boldsymbol\xi_b^{f}; P_k^*(\bu), f)\big),
\label{eq:nerf_bone}
\end{equation}
where $\boldsymbol\xi_b^{f}$ is a learnable latent code for encoding the $b^\text{th}$ bone at the $f^\text{th}$  frame; both $P_k^*(\bu)$ and $f$ are sent to the MLP as conditions to obtain ${\bf J}^{f}_b$. Thus $\bJ^{f}$ is also expected to be continuous \emph{w.r.t.} $P_k^*(\bu)$ and $f$.

Let $k$ index over intersected Gaussian surfels along the  camera ray that emanates from the frame pixel $\bar\bx^f$. Denote the  normal of the $k^\text{th}$ intersected  surfel as $\bn_k(\bar\bx^{f})$, and the surface normal at the interaction of the object surface with the camera ray as $\bN(\bar\bx^{f})$ which is  estimated by the nearby depth point, denoted as  $\bp^f$. Similar to the normal regularization~\cite{Huang2DGS2024},  $\bN(\bar\bx^{f})$ is computed with finite differences using the nearby depth point $\bp^{f}$ at the warped state frame $f$, and  $\bn_k(\bar\bx^{f})$ is computed by aligning its value with the surface normal with an added loss function $\mathcal{L}_\mathrm{N}$, 
\begin{equation}
\bN(\bar\bx^{f})= \frac{\nabla_x \bp^f \times \nabla_y \bp^f}{|\nabla_x \bp^f \times \nabla_y \bp^f|}, \quad \mathcal{L}_\mathrm{N}=\sum_{k} \omega_k (1-\bn_k^{\top}\bN),
\label{eq:normal}
\end{equation}
where $\omega_k = \alpha_k\,{\cG}_k(\bu(\bar\bx^{f}))\prod_{j=1}^{k-1} (1 - \alpha_j\,{\cG}_j(\bu(\bar\bx^{f})))$ denotes the blending weight of the $k^\text{th}$ intersected Gaussian surfel.

In summary, by optimizing the continuous warping field and aligning the surfel normals with the estimated surface normals in the warped state, we encourage that all Gaussian surfels locally approximate the actual object surface without large disruption from non-rigid deformation and frame distortion.
\begin{figure*}[t]
    \centering
     \hskip-0.05in
\includegraphics[width=1.73\columnwidth]{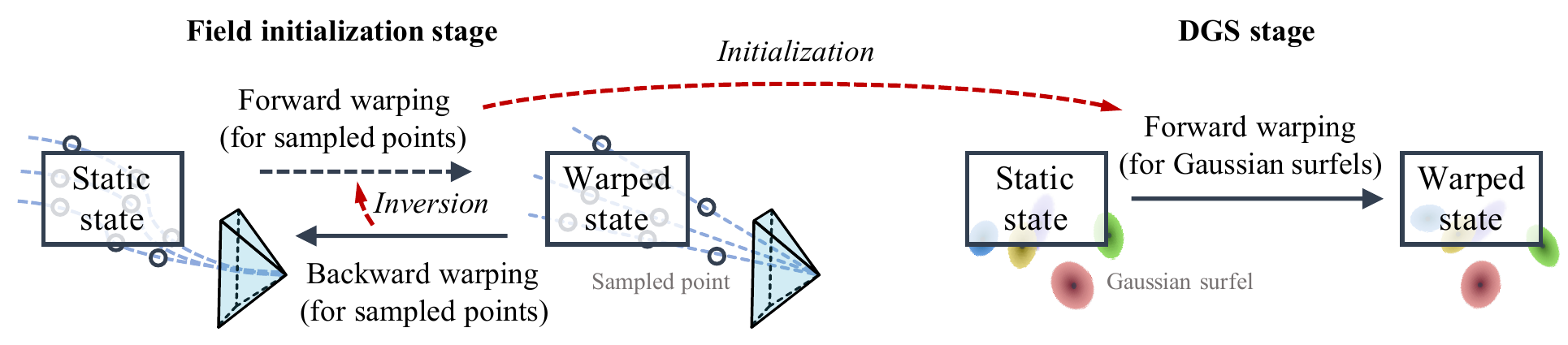}
    \vskip-0.01in
\caption{Illustration of the pipeline of Video4DGen, including the initialization stage and the DGS stage. }   \label{fig:Video4DGen}\vskip-0.04in
\end{figure*}

\textbf{Addressing appearance inconsistency.} 
\label{appearance_inconsistency} When Gaussian surfels are well reconstructed, their normal tends to be aligned with the surface normal which makes the gradient of density along the surface normal very large,
\begin{align}
\frac{d\cG(\bu) }{dx}&=-\mathcal{H}({\eta}, \gamma)x\exp\left(-\frac{\mathcal{H}(\eta, \gamma)x^2}{2}\right),
\label{eq:normal_large}
\end{align}
where $\cG(\bu)= \exp\left(-\frac{u^2+v^2}{2}\right)$ is the density of a surfel, and $x$ goes along the surface normal; $\mathcal{H}(\eta, \gamma)=\frac{1}{\sin^2(\eta)}+\frac{1}{\sin^2(\gamma)}$ with $\eta$ and $\gamma$ being the angles between $u$, $v$, and the surface, respectively.  Considering that $\eta$ and $\gamma$ are very small, the density is sensitive to the motion along the surface normal. Consequently, the texture flickering could be caused by Gaussian surfels moving back and forth due to the gradient direction, changing their depth order over time. When the front-back relationship between the rear surfels and surface surfels shifts, the texture flickering occurs accordingly.

To alleviate this texture flickering for a fine-grained appearance, we introduce a refinement process that elevates surfels to Gaussian ellipsoids, providing a more robust representation during warping. Specifically, we learn refinement terms for adjusting the rotation matrices $\bR_k^{*}$ and scaling matrices $\bS_k^{*}$ (defined in Eq.~(\ref{eq:2dgaussian})) in the static state. 
We suppose the refinement terms  are $\Delta\bR_k^{*}\in\SO(3)$ and $\Delta\bS_k^{*}\in\mathbb{R}^{3\times 3}$, respectively. Note that the third-axis of $\Delta\bS_k^{*}$ is no longer necessarily $0$. During refinement, we remain the center points $\bp_k^*$ and the warping  $\bJ^{f}$ (\emph{i.e.}, including both $\tilde\bR^{f}$ and $\tilde\bT^{f}$) to be unchanged. The new warped process is,
\begin{align}\hskip-0.1in
P_k^{\prime f}(\bu)=\begin{bmatrix}
        \tilde\bR^{f}(\Delta\bR_k^{*}\bR_k^{*})(\bS_k^{*}+\Delta\bS_k^{*}) & \hskip-0.035in \tilde\bR^{f}\bp_k^*+\tilde\bT^{f} \\
    \end{bmatrix}(u,v,1,1)^\top\hskip-0.035in.
    \label{eq:warping_refine}
\end{align}

During the training of DGS, we maintain two branches including one with refinement and one without. In the warped state, both branches are jointly trained with shared warping functions and centers of Gaussian primitives\footnote{Here, since the third-axis of the refined scaling matrix is not necessarily 0,  we adopt ``Gaussian primitive'' for commonly referring to both Gaussian surfel and the refined Gaussian.}. Due to the involvement of $\Delta\bR_k^{*}$ and $\Delta\bS_k^{*}$, both branches have different rotation and scaling matrices of Gaussian primitives.

\textbf{Rasterization.} For a frame pixel $\bar\bx$ and its corresponding camera ray, we first compute the intersection coordinates with Gaussian primitives along the ray using the static-state methods~\cite{kerbl3Dgaussians,Huang2DGS2024}.  We then obtain warped-state intersection coordinates based on Eq.~(\ref{eq:warping}) and  Eq.~(\ref{eq:warping_refine}). Finally, we  perform  volume rendering process~\cite{Huang2DGS2024} to integrate alpha-weighted appearance along the ray 
\begin{equation}
\br(\bar\bx^f) = \sum_{k} \bc_k\,\alpha_k\,\cG_k\big(\bu(\bar\bx^f)\big) \prod_{j=1}^{k-1} \big(1 - \alpha_j\,\cG_j\big(\bu(\bar\bx^f)\big)\big),
\label{eq:2dgs}
\end{equation}
where  $k$ indexes over intersected Gaussian primitives along the ray that emanates from the frame pixel $\bar\bx^f$; $\alpha_k$ and $\bc_k$ denote the opacity and view-dependent appearance parameterized with spherical harmonics of the $k^\text{th}$ Gaussian surfel, respectively; $\cG_k(\bu(\bar\bx^f)) = \exp\left(-\frac{u^2+v^2}{2}\right)$  corresponds to the $k^\text{th}$ intersection point $\bu(\bar\bx^f)$ which could be directly calculated when given $P_k^{f}(\bu)$ or $P_k^{\prime f}(\bu)$ and the corresponding local coordinate system. During implementation, $\cG_k(\bu(\bar\bx^f)))$ is further applied a low-pass filter following~\cite{botsch2005high,Huang2DGS2024}.

A detailed architecture of DGS is depicted in Fig.~\ref{fig:method}. Important symbols are summarized in the Appendix.

In summary, generated videos often exhibit more unexpected large-scale movements (non-rigidity and distortion) and small-scale anomalies (flickering and float-occlusion) compared to real videos. To address these challenges, we design  DGS   in a coarse-to-fine manner. The coarse part contains time-varying warpings, $\tilde\bR^{f}$ and $\tilde\bT^{f}$ in Eq.~(\ref{eq:warping}), to model the basic movement and register the camera and root pose. We also employ motion regularization in Eq.~(\ref{eq:nerf_bone}) and  field initialization (Sec.~\ref{subsec:Video4DGen}) to handle large-scale non-rigid movements and distortions even with limited viewpoints. Compared with  applying Gaussian primitives with dense motion, our sparse-bone design  alleviates overfitting \emph{w.r.t.} motions. The refinement part uses a time-invariant rotation $\Delta\bR_k^{*}$ and scaling $\Delta\bS_k^{*}$ in Eq.~(\ref{eq:warping_refine}) to alleviate overfitting \emph{w.r.t.} flickering and float-occlusion. Verification results will be shown in Sec.~\ref{subsec:ablation}.

\subsection{Field Initialization}
\label{subsec:Video4DGen}
Given that the camera trajectory of generated videos is unknown, SfM methods like COLMAP~\cite{schonberger2016structure,DBLP:conf/eccv/SchonbergerZFP16} struggle to converge due to rigidity violations. Additionally, since the background of generated videos appears to exhibit soft deformation or flickering colors, proper estimation of camera/subject poses through background SfM is hindered. Therefore, one of the primary challenges of 4D representation based on generated videos is the initialization of camera/subject poses and  subject motion. And preserving temporal consistency in texture and geometry is tough which  complicates the process of camera registration~\cite{DBLP:conf/cvpr/YangVNRVJ22}.

To address this, we design an implicit field before performing DGS to initialize the camera poses and establish the continuous warping field in Eq.~(\ref{eq:nerf_bone}). In this part, we propose the \textbf{field initialization} as another key component of our pipeline to initialize the continuous warping field of DGS for fast and stable convergence, as depicted in Fig.~\ref{fig:Video4DGen}, with details described below.

{We first train a neural Signed Distance Function (SDF)~\cite{DBLP:conf/nips/WangLLTKW21}, leveraging the same warping structure with bones as utilized in  DGS. While DGS transforms Gaussian surfels from the static state to the warped state for rasterization, the neural SDF maps points along camera rays from the warped state back to the static state.  For optimization ease, we define a root pose $\bG^{f}\in\SE(3)$  representing global transformation of the subject or scene, such as the per-frame facing direction of the phoenix case in Fig.~\ref{fig:intro-image}~(h).  This root pose $\bG^{f}$ is shared across all sample points of the neural SDF at frame $f$, and the camera pose optimization is absorbed into the optimization of $\bG^{f}$~\cite{DBLP:conf/cvpr/YangVNRVJ22,wang2023rpd}.  Specifically, $\mathbf{X}^{f}=\bG^{f}\mathbf{J}^{f} \mathbf{X}^{*}$, where $\mathbf{X}^{*}$ is a sample point in the static state for querying the neural SDF, and $\mathbf{X}^{f}$ is its  corresponding point at frame $f$.}

{Similarly, DGS could be enhanced by incorporating $\bG^{f}$ into Eq.~(\ref{eq:warping}). Both the neural SDF and DGS share the same design for  transformations  $\bJ^{f}$ and $\bG^{f}$. Optimizing these transformations in the neural SDF initializes those used in DGS. Further details and distinctions of $\bJ^{f}$ and $\bG^{f}$ will be provided in Sec.~\ref{sec:4dgen_multi}, specifically in Eq.~(\ref{eq:warpingJ}) and Eq.~(\ref{eq:warpingG}), within the context of multi-video training.}

During the rendering of the neural SDF, we perform backward warping on the warped-state sample points to the static state, 
\begin{equation}
\mathbf{J}^{f, -1} = \mathcal{R}\Big(\sum_{b=1}^{B} w_{b}^{f} \mathcal{Q}(\mathbf{J}^{f}_{b})^{-1}\Big),\quad {\mathbf{X}^{*} = \mathbf{J}^{f, -1}(\mathbf{G}^{f})^{-1} \mathbf{X}^{f},}
\label{eq:backward_warping}
\end{equation}
{where the former equation follows the  inverse format of Eq.~(\ref{eq:warpingJ1}). By querying the SDF with a sample point $\mathbf{X}^{*}$  in the static state, we render RGB and compute the photometric loss to optimize the SDF and the warping field. Subsequently, we use network weights learned by the neural SDF to initialize $\mathbf{MLP}(\cdot)$ in Eq.~(\ref{eq:nerf_bone}).}

{Nevertheless, there are two main discrepancies between the  neural SDF warping and the  DGS warping. First,   sample points for the neural SDF are spread throughout the camera frustum, while those for later DGS are located on the object surface. Second, for the neural SDF, we train the warping from the warped state back to the static state, whereas DGS uses  forward warping. To address these differences and ensure accurate forward warping, we incorporate a cycle loss~\cite{DBLP:journals/corr/abs-2206-15258} to deduce the common forward warping from the neural SDF's backward warping, }
\begin{equation}
{\mathcal{L}_{\mathrm{cyc}} = \big\| \bG^{f}\mathbf{J}^{f} \mathbf{X}^{*} - \mathbf{X}^{f} \big\|_2^{2},}
\label{eq:cycle_loss}
\end{equation}
{where $\mathbf{X}^{f}$ is randomly from the camera ray required by the neural SDF training and the object surface to reduce the gap for DGS.}

After initialization, we extract the canonical space mesh using marching cubes and initialize Gaussian surfels on it. The $0^\text{th}$ order spherical harmonics are set to the RGB values of the closest vertices. The warping field and learned camera poses are retained.

By integrating DGS with field initialization, Video4DGen can achieve high-fidelity 4D generation from a single generated video.

\subsection{4D Generation from Multiple Generated Videos}
\label{sec:4dgen_multi}
In this part, we detail the pipeline to generate a group of 4D contents (represented by DGS) based on multiple generated videos, namely, $M$ could be larger than $1$, as detailed below. 

\textbf{Multi-video DGS alignment.} Given a single input image,  we generate various videos  by guiding the movement of the main subject (could be either object or character) in the image using different text descriptions. Each video, denoted as $\Vv^m \in \mathbb{R}^{F \times H \times W \times 3}$ where $m=1,\text{\mycdots},M$, is generated through an image-to-video process.  Since each set of videos begins with the same initial frame, we suppose that the foundational Gaussian surfels from this first frame are common across all videos. These surfels are then individually transformed into a 4D representation for each video sequence. We also propose that identical Gaussian surfels correspond to the same semantic point throughout all videos and across every frame. The alignment within each video is facilitated by utilizing DINO features~\cite{oquab2023dinov2} to identify key feature  patterns.

\textbf{Static-state sharing and continuous root pose optimization.} In this part, we introduce two techniques. Firstly, to guarantee the consistency in the appearance and geometric structure of the subject, we maintain a set of Gaussian surfels $\{P_k^*(\bu)\}_{k=1}^K$ in the static state that is shared across various frames and multiple videos. Secondly, we optimize the per-frame root pose, which refers to the fundamental orientation and position of the subject in each frame. The root pose, denoted as $\bG^{f,m}$ belonging to the special Euclidean group $\SE(3)$, represents the learned orientation of the subject and is irrelevant to the index $k$.

Specifically, the $k^\text{th}$ Gaussian surfel in the static state is transformed to its warped state at the $f^\text{th}$ frame of the $m^\text{th}$ video through a learned transformation $\bJ^{f,m}$ and the root pose $\bG^{f,m}$, both belonging to $\SE(3)$. The deformation of the $k^\text{th}$ Gaussian surfel could be computed by
\begin{align}
P_k^{f,m}(\bu)=\bG^{f,m}\bJ^{f,m}P_k^*(\bu),
    \label{eq:warping_root}
\end{align}
where $P_k^*(\bu)$ denotes the coordinate in the world space of $\bu=(u,v)$ taking $\bo_k^*$ as the coordinate origin, as previously introduced in Sec.~\ref{subsec:modeling}. $P_k^{f,m}(\bu)$ is the warped coordinate in the world space.

We use MLPs to optimize  continuous fields $\bJ^{f,m}$ and $\bG^{f,m}$ for the $m^\text{th}$ video, represented by $\mathbf{MLP}_\mathrm{J}^m(\cdot)$ and $\mathbf{MLP}_\mathrm{G}^m(\cdot)$, respectively. The per-surfel transformation $\bJ^{f,m}$ is expected to vary smoothly \emph{w.r.t.} both the coordinate $P_k^*(\bu)$ and the  frame index $f$. In contrast, the root pose $\bG^{f,m}$ changes  continuously \emph{w.r.t.} only the frame index $f$ and is shared across all Gaussian surfels within the same frame. The optimization is represented by
\begin{align}
\bJ^{f,m}&=\mathcal{R}\Big(\sum_{b=1}^{B} w_{b}^{f,m} \mathbf{MLP}_\mathrm{J}^m\big(\boldsymbol\xi_b^{f,m}; P_k^*(\bu), f\big)\Big),\label{eq:warpingJ}\\
\bG^{f,m}&= \mathcal{R}\Big(\mathbf{MLP}_\mathrm{G}^m\big(\boldsymbol\zeta^{f,m}; f\big)\Big),
\label{eq:warpingG}
\end{align}
where in Eq.~(\ref{eq:warpingJ}) we apply the dual quaternion skinning technique~\cite{DBLP:conf/si3d/KavanCZO07} with $B$ being the number of bones and $w_{b}^{f,m}$  being the $b^\text{th}$ weighting factor. The operation $\mathcal{R}(\cdot)$ in  Eq.~(\ref{eq:warpingJ}) and Eq.~(\ref{eq:warpingG}) refers to the inverse quaternion process, ensuring that the output lies within $\SE(3)$. $\boldsymbol\xi_b^{f,m}$ and $\boldsymbol\zeta^{f,m}$ are learnable latent codes.

\textbf{Pose-guided frame and video sampling.} As previously mentioned in the multi-video generation, for each case, we have a subject surrounding video ($\Vv^0$) and also $M$ videos capturing subject movements, resulting in $M$ corresponding 4D representations. Incorporating multiple videos increases the likelihood of capturing a broader range of root poses of the subject, which in turn enriches the dynamics and details of the generated 4D contents.

During the DGS optimization in Eq.~(\ref{eq:overall_dgs}), a video and a frame are sampled for each training step. However, since 4D representation involves learning the spectrum range of 360$^\circ$ root poses and motions, uniformly sampling frames and videos  causes the model to disproportionately focus on a narrow range of root poses while neglecting others. This results in a limited set of reconstructed poses and could lead to overfitting during DGS optimization.

Therefore, instead of directly sampling indexes of frames or videos, we  design a (root) pose-guided sampling strategy based on the per-frame/video root pose $\bG^{f,m}$. Since $\bG^{f,m}$ belongs to $\SE(3)$ and could also be decomposed to  both rotation and translation components. Denote the rotation matrix  as $\hat\bR^{f,m}\in\SO(3)$ which is the first $3\times3$ sub-matrix of $\bG^{f,m}$. We can extract the rotation angle $\phi^{f,m}$ (in degrees) from the rotation matrix by 
\begin{align}
\phi^{f,m} = \frac{180}{\pi} \cdot \cos^{-1} \left( \frac{\mathrm{Tr}(\hat\bR^{f,m}) - 1}{2} \right),
\end{align}
where $\mathrm{Tr}(\hat\bR^{f,m})$ is the trace of the rotation matrix $\hat\bR^{f,m}$.

During the sampling process, we first sample a rotation angle $\phi_\text{sampled}$  uniformly from the range $[0,360]$, and then identify the frame index $f$ and the video index $m$ of which the rotation angle $\phi^{f,m} $ is  closest to $\phi_\text{sampled}$, namely, 
\begin{align}
\{f_{\text{sampled}},m_{\text{sampled}}\} =\mathop{\arg\min}\limits_{f,m}  |\phi^{f,m}  - \phi_\text{sampled}|,
\end{align}
where $f$ and $m$ in ranges $[1,\text{\myldots},F]$ and  $[1,\text{\myldots},M]$, respectively. We omit the subscript ``$\text{sampled}$'' for simplicity in subsequent sections. 

\subsection{Novel-view Video  Generation with 4D Guidance}
\label{sec:vidgen}

Based on the multi-video generated 4D content (DGS) described in Sec.~\ref{sec:4dgen_multi}, we now focus on generating novel-view videos. This process leverages  DGS as intermediate guidance, which can be manipulated to influence the final output.

When given a novel viewpoint, the rendering results of DGS might exhibit blurry rendering in unobserved areas, thus  attempting to generate videos directly under DGS conditions leads to erratic video content in those specific regions. 
Therefore, our initial approach is to identify  the problematic regions, subsequently creating a confidence map. This map is  utilized to direct the video generation process, ensuring a more controllable video outcome.

\textbf{DGS filtering with confidence map.}  For a frame pixel $\bar\bx^{f,m}$ at the $f^\text{th}$ frame of the $m^\text{th}$ video,  and a camera ray that emanates from $\bar\bx^{f,m}$, we could compute the normal of the $k^\text{th}$ intersected  surfel  $\bn_k(\bar\bx^{f,m})$ and the surface normal  $\bN(\bar\bx^{f,m})$ by substituting $\bx^{f}$ in Eq.~(\ref{eq:normal}) with $\bx^{f,m}$.

We design an alignment error of the pixel $\bar\bx^{f,m}$, denoted as $e(\bar\bx^{f,m})\in\mathbb{R}$ by leveraging the distance of both normal values,
\begin{equation}
e(\bar\bx^{f,m})=\Big\|\sum_{k} \omega_k\bn_k(\bar\bx^{f,m})-\bN(\bar\bx^{f,m})\Big\|_2^2.
\end{equation}

To guide the video generation process with DGS, we aim for accurate rendering with each pixel exhibiting a low alignment error. Suppose for the $m^\text{th}$ video, the confidence map is  represented as $\mathcal{M}^m\in\{0,1\}^{ F \times   H \times   W}$. The value of  $\mathcal{M}^m$ at each pixel $\bar\bx^{f,m}$ is obtained by comparing $e(\bar\bx^{f,m})$ with a pre-defined threshold $h$,
\begin{equation}
\mathcal{M}^m(\bar\bx^{f,m})=\mathbb{I}_{\{e(\bar\bx^{f,m})<h\}},
\end{equation}
where $\mathbb{I}_{\{\cdot\}}$ denotes the indicator function, assigning a value of $1$ when the specified condition is met, and $0$ otherwise.

Hence, for the $m^\text{th}$ video, we leverage $\mathcal{M}^m$ to determine how the rasterized result $\br^m$ as mentioned in Eq.~(\ref{eq:2dgs}) is used to guide the video generation, which is detailed below.

\begin{figure*}[t]
    \centering  \hskip-0.06in
\includegraphics[width=1.9\columnwidth]{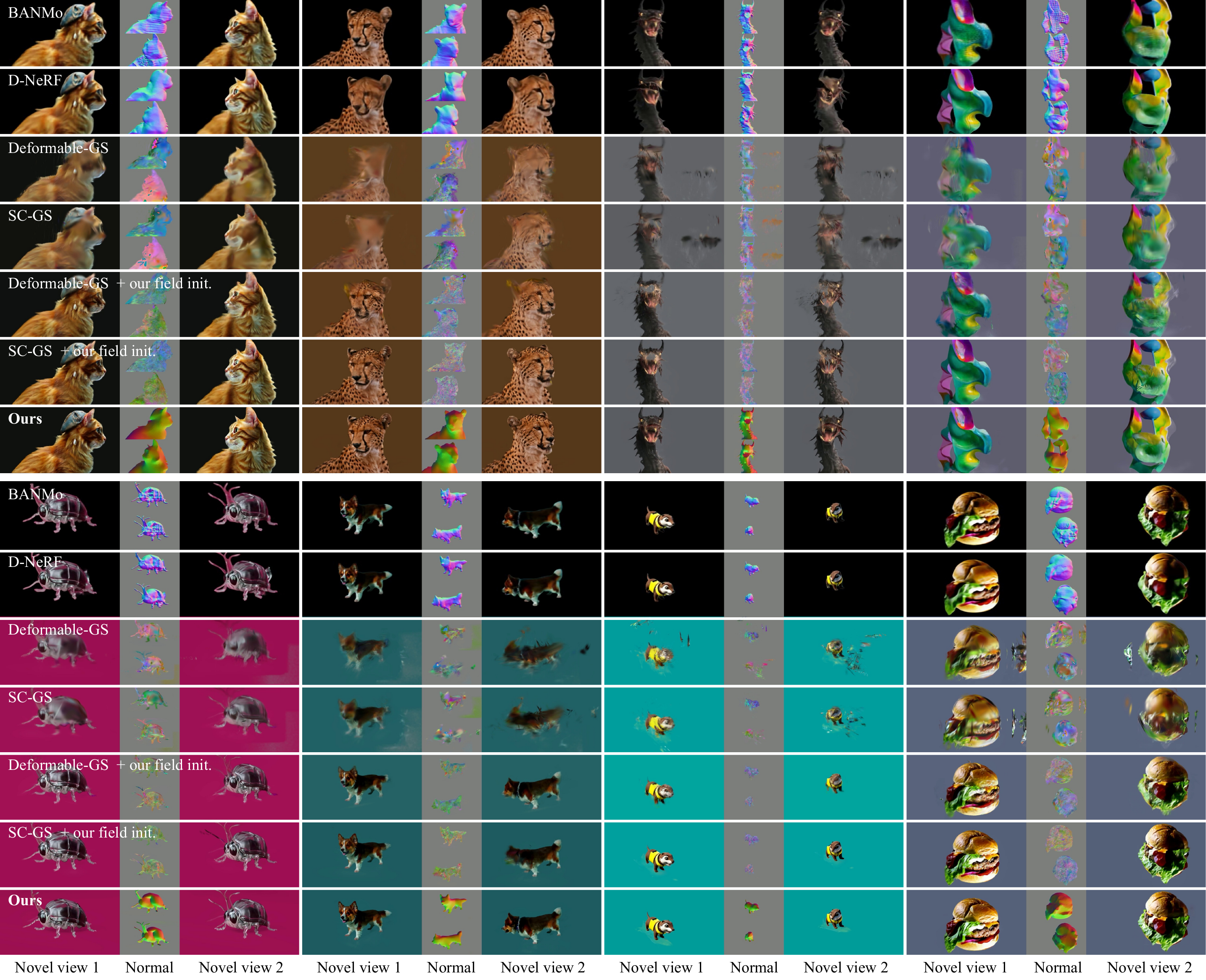}
    \vskip-0.03in
    \caption{Novel-view qualitative evaluation compared with state-of-the-art methods including NeRF-based methods (BANMo~\cite{DBLP:conf/cvpr/YangVNRVJ22} and D-NeRF~\cite{pumarola2021d}) and Gaussian splatting-based methods (Deformable-GS~\cite{yang2023deformable3dgs} and SC-GS~\cite{huang2023sc}). We also apply our field initialization (Sec.~\ref{subsec:Video4DGen}) to baseline approaches for a fair comparison, and these variants are denoted as ``+ our field init.''. Best view in color and zoom in.}
    \label{fig:qualitative}
\end{figure*}

\textbf{Novel-view video generation given filtered DGS.} We adhere to the same diffusion time step schedule $0=\tau_0 < \tau_1<\text{\mycdots}<\tau_{S}=T$ for two types of regions: the masked regions, which include pixels $\bar\bx^{f,m}$ where $\mathcal{M}^m(\bar\bx^{f,m})=1$, and the unmasked regions for all other pixels. Both types of regions begin in a state of noise and gradually become clearer as the time step decreases.

For the $m^\text{th}$ video, we handle the masked regions by applying a transformation to the rasterized result $\br^m$. This transformation aims to gradually move the data distribution towards a standard normal distribution. Suppose $\y_{\tau_t}^m$ is the intermediate representation of the data, $\text{Enc}(\cdot)$ is the encoder function, and $\bm{\epsilon}$ is the noise vector as introduced in Sec.~\ref{subsec:preliminary_video}. The process is described by 
\begin{equation}
\label{eq:exchange-bn}
\hskip-0.2em
\y_{\tau_t}^m =
\begin{cases}
(1-\tau_t)\text{Enc}(\br^m)+\tau_t\bm{\epsilon},   \quad\quad \quad\quad\;\;\;\;\,\text{RF} \\
\text{Enc}(\br^m)+\exp F_\mathcal{N}^{-1}(\tau_t|P_m,P_s^2)\bm{\epsilon},\; \;\,\text{EDM}
\\
a_{\tau_t}\text{Enc}(\br^m)+b_{\tau_t}\bm{\epsilon}, \quad    \quad\quad\quad\quad\quad\;\;\,\;\text{VP}
\end{cases}
\end{equation}
where we provide cases of common flow trajectories for the video diffusion model, including Rectified Flow (RF)~\cite{DBLP:conf/iclr/LiuG023}, EDM~\cite{DBLP:conf/nips/KarrasAAL22}, and  Variance Preserving (VP)~\cite{DBLP:conf/iclr/0011SKKEP21}. Here, $F_\mathcal{N}^{-1}$ is the quantile function of the normal distribution with mean $P_m$ and variance $P_s^2$ (see~\cite{DBLP:conf/nips/KarrasAAL22} for more details), and $a_{\tau_t}^2+b_{\tau_t}^2=1$.

In the unmasked regions, we perform denoising steps using the predicted velocity $\bv_\Theta(\z_{\tau_{t}}^m,\tau_{t})$, updating the latent state as follows,
\begin{equation}
\tilde\z_{\tau_{t-1}}^m=\z_{\tau_{t}}^m+\bv_\Theta(\z_{\tau_{t}}^m,\tau_{t})(\tau_{t}-\tau_{t-1}).
\end{equation}

Then the final latent state $\z_{\tau_{t-1}}$ is  computed by combining  the denoised unmasked regions and the transformed masked regions,\begin{equation}
\z_{\tau_{t-1}}^m=(1-\mathcal{M}^m)\tilde\z_{\tau_{t-1}}^m+\mathcal{M}^m\y_{\tau_{t-1}}^m.
\end{equation}

The  $m^\text{th}$  video with 4D guidance is obtained by $\text{Dec}(\z_0^m)$ where $\text{Dec}(\cdot)$ is the video decoder.

By these designs, we could perform 4D-guided video generation, and one special case is the multi-camera video generation which generates consistent videos by capturing the same motion sequence at multiple camera perspectives. We also provide ablation studies to verify the effectiveness of using confidence map.

Leveraging its capacity for 4D-guided novel-view video generation, Video4DGen could generate a 360$^\circ$ rendering of the subject, which in turn improves the 4D generation process across a wide range of camera angles. This establishes Video4DGen as a mutual optimization framework for both 4D and video generation.

\begin{figure*}[t]
   \centering
\includegraphics[width=1.8\columnwidth]{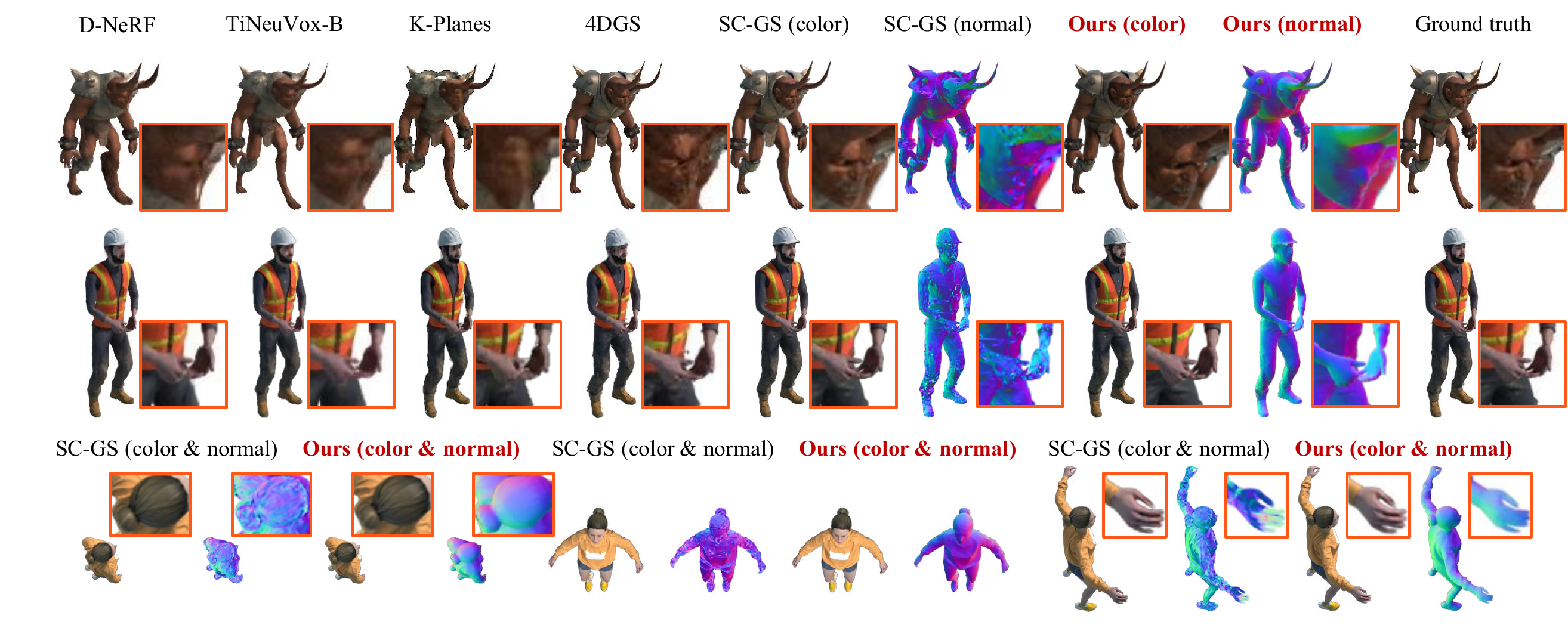}
   \vskip-0.07in
   \caption{Qualitative object-level 4D results on  \textbf{D-NeRF}  dataset. Best view in color and zoom in.}
   \label{fig:dnerf}
\end{figure*}

\begin{table*}[t]
\centering
\caption{Novel-view quantitative results on generated videos. Evaluation metrics are PSNR, SSIM, and LPIPS. We report results on three single videos and the averaged results over 30 single videos.}
\tablestyle{1pt}{1}
\resizebox{2\columnwidth}{!}{
\begin{tabular}{@{}l|ccc|ccc|ccc|ccc@{}}
\toprule[1pt]
\multirow{2}{*}{Method} & \multicolumn{3}{c@{}|}{Cat} & \multicolumn{3}{c@{}|}{Cheetah} & \multicolumn{3}{c@{}|}{Dragon} & \multicolumn{3}{c@{}}{Average over 30 videos}\\ 
& PSNR~$\uparrow$ & SSIM~$\uparrow$ & LPIPS~$\downarrow$ & PSNR~$\uparrow$ & 
SSIM~$\uparrow$ & LPIPS~$\downarrow$ & PSNR~$\uparrow$ & SSIM~$\uparrow$ & LPIPS~$\downarrow$ & PSNR~$\uparrow$ & SSIM~$\uparrow$ & LPIPS~$\downarrow$ \\
\midrule
BANMo~\cite{DBLP:conf/cvpr/YangVNRVJ22} & 15.10 & 0.6514 & 0.2575 & 13.15 & 0.5921 & 0.3241 & 18.48 & 0.6423 & 0.3500 & 13.62 $\pm$ 2.99 & 0.6153 $\pm$ 0.0714 & 0.3738 $\pm$ 0.0665 \\

D-NeRF~\cite{pumarola2021d} & 15.15 & 0.6537 & 0.2657 & 13.21 & 0.5930 & 0.3344 & 18.53 & 0.6489 & 0.3527 & 21.01 $\pm$ 2.86 & 0.8519 $\pm$ 0.0717 & 0.1522 $\pm$ 0.0754 \\

\midrule
Deformable-GS~\cite{yang2023deformable3dgs} & 19.09 & 0.7815 & 0.2434 & 20.35 & 0.8039 & 0.1982 & 24.19 & 0.9100 & 0.0992 & 13.22 $\pm$ 3.42 & 0.5934 $\pm$ 0.0535 & 0.3749 $\pm$ 0.0763 \\

SC-GS~\cite{huang2023sc} & 19.46 & 0.7867 & 0.2405 & 20.87 & 0.8123 & 0.1919 & 24.03 & 0.9083 & 0.1009 & 21.17 $\pm$ 2.69 & 0.8547 $\pm$ 0.0691 & 0.1504 $\pm$ 0.0737 \\

Deformable-GS + our field init. & 21.94 & 0.8123 & 0.1816 & 22.41 & 0.8200 & 0.1687 & 26.05 & 0.9218 & 0.0894 & 22.63 $\pm$ 2.14 & 0.8469 $\pm$ 0.0438 & 0.1452 $\pm$ 0.0354 \\

SC-GS + our field init. & 23.25 & 0.8268 & 0.1574 & 23.70 & 0.8338 & 0.1497 & 28.40 & 0.9375 & 0.0686 & 24.75 $\pm$ 2.11 & 0.8680 $\pm$ 0.0440 & 0.1201 $\pm$ 0.0359 \\

\textbf{Ours} & \best\textbf{24.63} & \best\textbf{0.8432} & \best\textbf{0.1559} & \best\textbf{25.68} & \best\textbf{0.8843} & \best\textbf{0.1117} & \best\textbf{28.58} & \best\textbf{0.9392} & \best\textbf{0.0618} & \textbf{27.30} $\pm$ \best\textbf{2.66} & \best\textbf{0.9152} $\pm$ \textbf{0.0602} & \best\textbf{0.0877}$\pm$ \textbf{0.0564} \\
\bottomrule[1pt]
\end{tabular}
}
\label{tab:main_results}
\end{table*}

\begin{table}[t]
\caption{Quantitative results on the built \textbf{SORA} benchmark.}
\centering
\tablestyle{10pt}{1}
\resizebox{1\columnwidth}{!}{
\begin{tabular}{@{}l|ccc}
\toprule[1pt]
\multirow{2}{*}{Method}& \multicolumn{3}{c@{}}{SORA benchmark}\\ 
 & PSNR~$\uparrow$ & MS-SSIM~$\uparrow$ & LPIPS~$\downarrow$ \\
\midrule
4DGS~\cite{wu20234dgaussians} & 12.15	 & .5609 & .2926  \\
Deformable-GS~\cite{yang2023deformable3dgs}	 & 12.72 & .5773 & .2861 \\
SC-GS~\cite{huang2023sc} & 14.81	 & .5914  & .2420  \\
SpacetimeGaussians~\cite{DBLP:conf/cvpr/LiCLX24} &   13.24 & .5836& .2633 \\
\textbf{Ours} & \best\textbf{19.05} & \best\textbf{.7323} & \best\textbf{.1839} \\
\bottomrule[1pt]
\end{tabular}}\vskip-0.05in
\label{tab:sora_results}
\end{table}

\begin{table}[t]\vskip-0.01in
\caption{Quantitative results on \textbf{D-NeRF} data with GT  poses.}\vskip-0.015in
\centering
\tablestyle{15pt}{0.9}
\resizebox{1\columnwidth}{!}{
\begin{tabular}{@{}l|ccc}
\toprule[1pt]
\multirow{2}{*}{Method}  & \multicolumn{3}{c@{}}{D-NeRF dataset (with GT poses)}\\ 
 & PSNR~$\uparrow$ & SSIM~$\uparrow$ & LPIPS~$\downarrow$ \\
\midrule
D-NeRF~\cite{pumarola2021d} & 31.69 & .975 & .0575 \\
TiNeuVox-B~\cite{DBLP:conf/siggrapha/FangYWX00N022} & 33.76& .983 & .0441  \\
Tensor4D~\cite{DBLP:conf/cvpr/ShaoZTL0L23} & 27.62 & .947  & .0471  \\
K-Planes~\cite{DBLP:conf/cvpr/Fridovich-KeilM23}  &  32.32  & .973 & .0382 \\
4DGS~\cite{wu20234dgaussians} &  34.01 & .987 & .0316 \\
{4D-Rotor~\cite{DBLP:conf/siggraph/DuanWDHCC24}} &  34.79 & .986 & .0332 \\
FF-NVS~\cite{DBLP:conf/iccv/GuoSDCY0DZ023} &  33.73  & .979 & .0357 \\
{Deformable-GS~\cite{yang2023deformable3dgs}}&  40.50 & .992 & .0102 \\
SC-GS~\cite{huang2023sc} &  \sbest43.31 & \sbest{.997}& \sbest.0063 \\
\textbf{Ours} &  \best\textbf{43.86} &  \best\textbf{.998} &  \best\textbf{.0059} \\
\bottomrule[1pt]
\end{tabular}}\vskip-0.1in
\label{tab:dnerfgt}
\end{table}

\begin{table}[t]
\caption{Quantitative results on \textbf{D-NeRF}  data without GT poses.}\vskip-0.016in
\centering
\tablestyle{4pt}{1}
\resizebox{1\columnwidth}{!}{
\begin{tabular}{@{}l|ccc}
\toprule[1pt]
 \multirow{2}{*}{Method} & \multicolumn{3}{c@{}}{D-NeRF dataset (without GT poses)}\\ 
 & PSNR~$\uparrow$ & SSIM~$\uparrow$ & LPIPS~$\downarrow$ \\
\midrule
BANMo~\cite{DBLP:conf/cvpr/YangVNRVJ22} & 14.4558 $\pm$ 2.9197 & .6528 $\pm$ .0481 & .3458 $\pm$ .0336 \\
D-NeRF~\cite{pumarola2021d} & 14.5561 $\pm$ 3.2744 & .6599 $\pm$ .0528 & .3410 $\pm$ .0416 \\
Deformable-GS~\cite{yang2023deformable3dgs} & 19.4895 $\pm$ 2.9886 & .9153 $\pm$ .0411 & .0775 $\pm$ .0392 \\
SC-GS~\cite{huang2023sc} &  \sbest 20.0486 $\pm$ 2.6920 & \sbest .9170 $\pm$ .0399 & \sbest .0764 $\pm$ .0387 \\
\textbf{Ours} & \best\textbf{29.0624} $\pm$ \best\textbf{2.0656} & \textbf{.9622} $\pm$ \best\textbf{.0218} & \best\textbf{.0319} $\pm$ \textbf{.0177} \\
\bottomrule[1pt]
\end{tabular}}\vskip-0.02in
\label{tab:dnerfwogt}
\end{table}

\begin{table}[t]
\caption{Quantitative results on \textbf{Neural 3D Video} data.}
\centering
\tablestyle{0.4pt}{0.98}
\resizebox{1\columnwidth}{!}{
\begin{tabular}{@{}l|ccccc}
\toprule[1pt]
 \multirow{3}{*}{Method} & \multicolumn{5}{c@{}}{Neural 3D Video dataset}\\ 
 & PSNR~$\uparrow$ & DSSIM$_1$~$\downarrow$ & DSSIM$_2$~$\downarrow$ & LPIPS~$\downarrow$ & \makecell{FPS~$\uparrow$\\(reported / tested)} \\
\midrule
StreamRF~\cite{DBLP:conf/nips/0011SW0T22} & 28.26 & - & -  &- & 10.9 / -\\
NeRFPlayer~\cite{DBLP:journals/tvcg/SongCLCCYXG23} & 30.69& .034 & - &.111 & 0.05 / -  \\
HyperReel~\cite{DBLP:conf/cvpr/Attal0RZ0OK23} & 31.10 & .036  & - & .096  & 2 / - \\
K-Planes~\cite{DBLP:conf/cvpr/Fridovich-KeilM23} &  31.63  & .- & .018 & - & 0.3 / -\\
MixVoxels-L~\cite{DBLP:conf/iccv/WangTLTS023} &  31.34&-  & .017 & .096 & 37.7 / - \\
MixVoxels-X~\cite{DBLP:conf/iccv/WangTLTS023}&  31.73 & - & .015 & .064  &4.6 / -\\
{RealTime4DGS~\cite{DBLP:conf/iclr/YangYP024}} &  30.62 & .036 & .020& .109 &72.8 / 90.5\\
SpacetimeGaussians~\cite{DBLP:conf/cvpr/LiCLX24} &   \sbest32.05 & \sbest.026 & \best\textbf{.014}&  \sbest.044 & 140 / 191\\
\textbf{Ours} & \best\textbf{33.15} & \best\textbf{.023} & \best\textbf{.014} & \best\textbf{.037} & - / \best\textbf{206} \\
\bottomrule[1pt]
\end{tabular}}\vskip-0.1in
\label{tab:neural_3d_video}
\end{table}

\begin{figure*}[t]
   \centering\vskip-0.1in
\includegraphics[width=1.8\columnwidth]{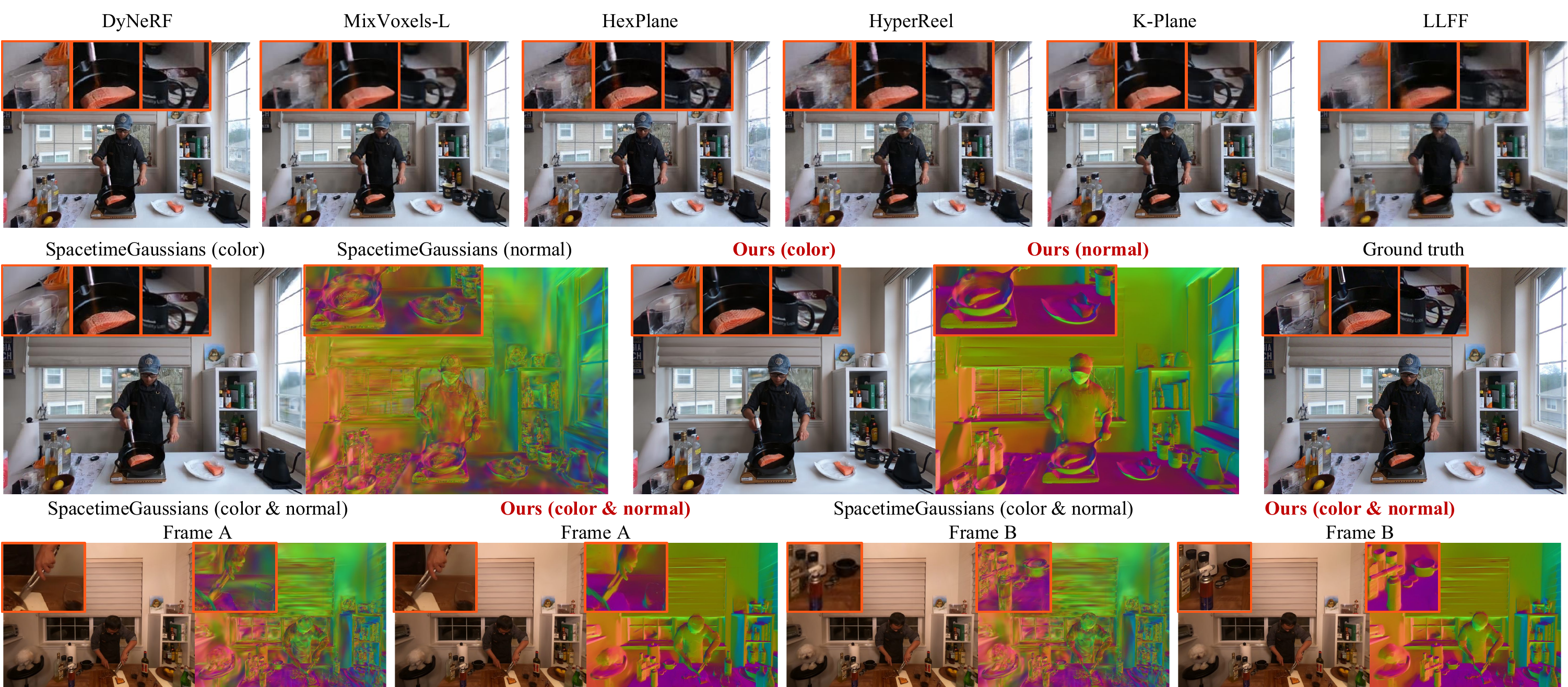}      \vskip-0.03in
\caption{Qualitative  real-scene 4D results on  \textbf{Neural 3D Video} dataset. Best view in color and zoom in.}
   \label{fig:neural3dvideo}
\end{figure*}
\vspace{-0.1in}

\begin{figure*}[t]
   \centering\vskip-0.1in
\includegraphics[width=1.8\columnwidth]{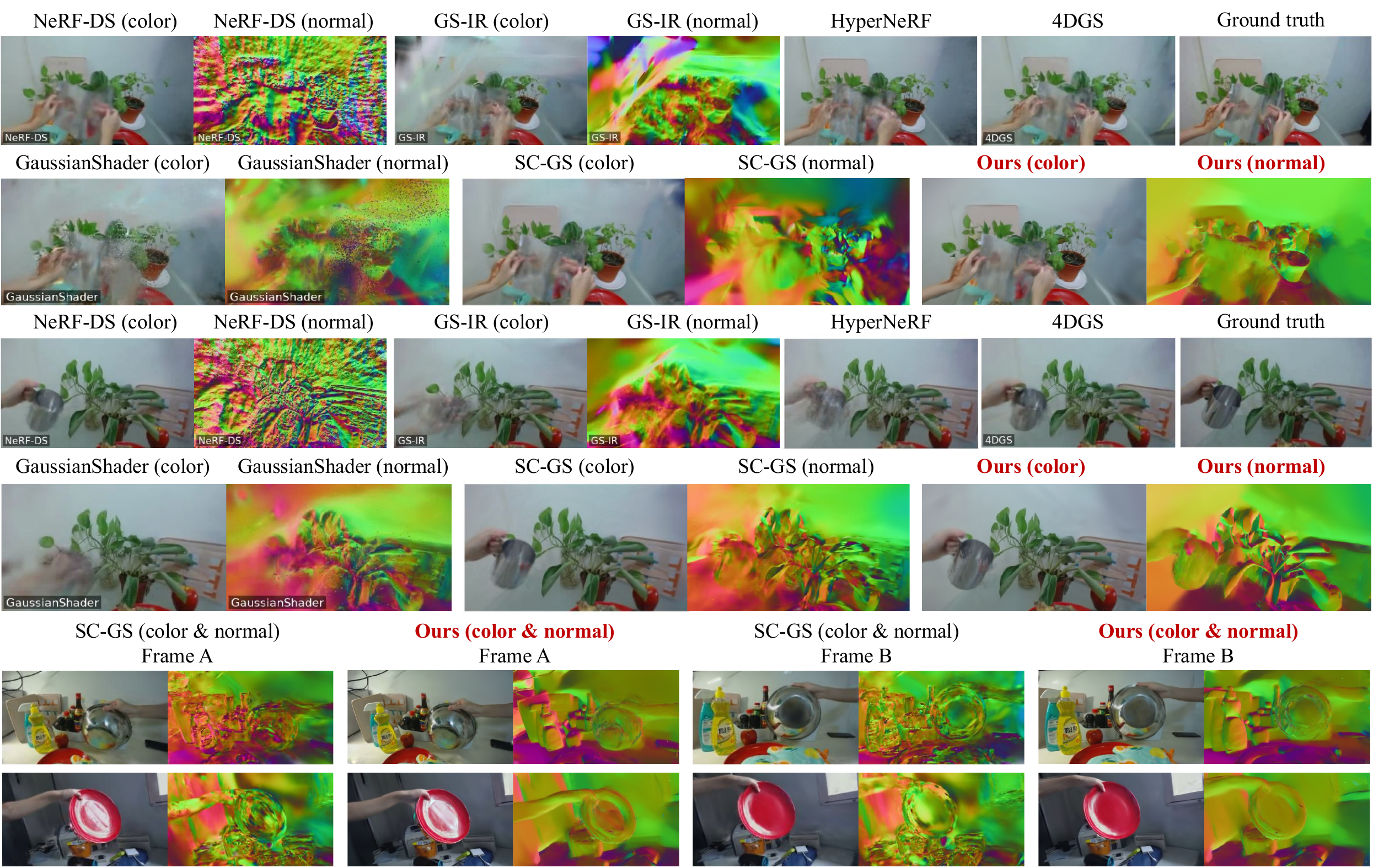}\vskip-0.03in
\caption{Qualitative real-scene 4D results on  \textbf{NeRF-DS} dataset. Best view in color and zoom in.}\vskip-0.07in
   \label{fig:method_nerfds}
\end{figure*}

\begin{figure*}[t]
   \centering
\includegraphics[width=1.8\columnwidth]{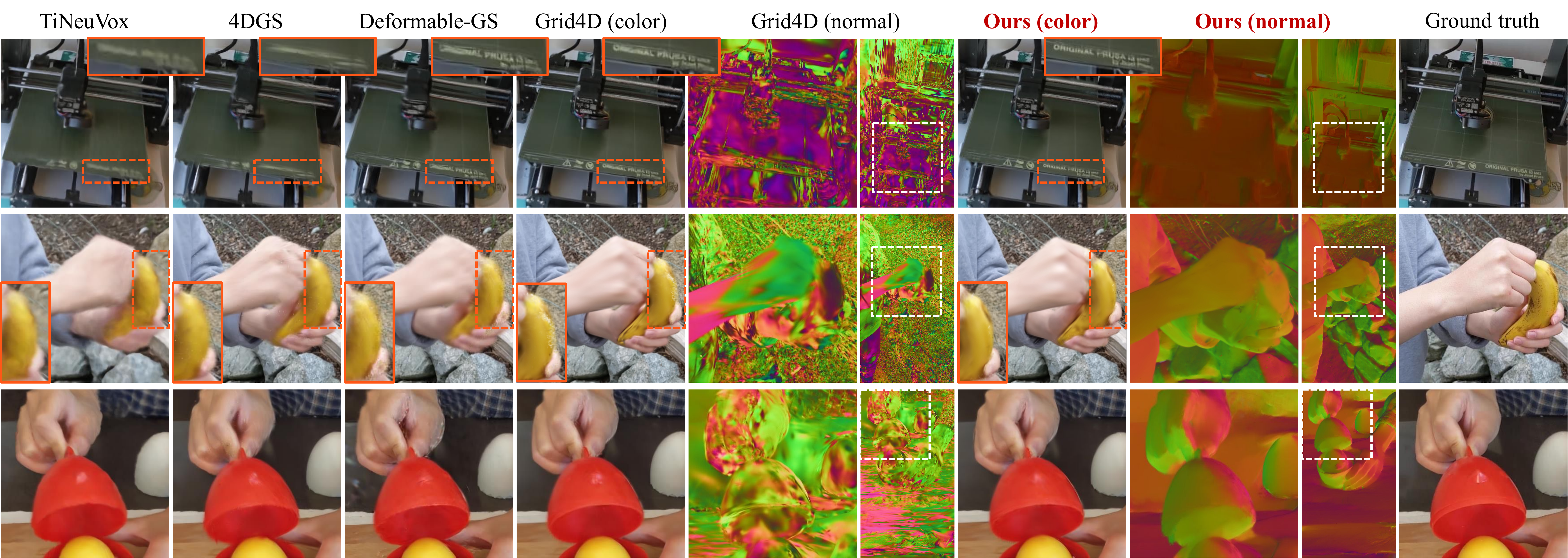}\vskip-0.03in
  \caption{Qualitative real-scene 4D results on  \textbf{HyperNeRF} dataset. Best view in color and zoom in.}\vskip-0.04in
   \label{fig:method_hypernerf}
\end{figure*}

\section{Experiment}
\label{sec:exp}
We provide an extensive evaluation by comparing both appearance and geometry against previous state-of-the-art  methods. We also analyze the contributions of each proposed component in detail.

\subsection{Implementation Details}
\label{subsec:Implementation}
{For the field initialization stage,} we use a NeRF-like~\cite{DBLP:conf/eccv/MildenhallSTBRN20} architecture with $8$ layers for volume rendering, and initialize MLP for predicting SDF as an approximate unit sphere~\cite{DBLP:conf/nips/YarivKMGABL20}. We obtain a neural SDF, a warping field, and camera poses after this stage. {For the DGS stage,}
we initialize centers of the Gaussian surfels with the sampled surface points extracted from the neural SDF, and initialize the warping field by the forward field from the first stage. The dimensions of the latent code embeddings  {$\boldsymbol\xi_b^{f,m}$ and $\boldsymbol\zeta^{f,m}$} are both set to $128$. Following BANMo~\cite{DBLP:conf/cvpr/YangVNRVJ22}, we adopt 25 bones to optimize skinning weights. For each case, the overall training takes over 1 hour on an Nvidia A800 GPU. Specifically, generating a 1080p video takes approximately 10 minutes. Preprocessing requires around 12 minutes, initialization takes another 10 minutes, and reconstruction takes 30 minutes. Rendering a 1080p (1920$\times$1080) image takes less than 0.1 second.

\subsection{Results of 4D Generation from Generated Videos}
A part of the qualitative results of our 4D generation is already shown in Fig.~\ref{fig:intro-image}. In this section, we  evaluate our method on generated videos and provide both qualitative and quantitative  results. We  strictly compare our method against existing state-of-the-art 4D reconstruction methods. Besides, since our focus is on the 4D reconstruction, to the best of our knowledge, there are no pose-free benchmarks for dynamic scenes. Thus we also build a new benchmark by collecting openly available videos from the SORA official webpage. For all the experiments conducted in this section, we follow the standard pipeline for dynamic reconstruction~\cite{DBLP:journals/tog/ParkSHBBGMS21,huang2023sc}, to construct our evaluation setup by selecting every fourth frame as a training frame and designating the middle frame between each pair of training frames as a validation frame.

\textbf{Qualitative evaluation.} We visually compare our DGS reconstructions with those from other state-of-the-art models, as illustrated in Fig.~\ref{fig:qualitative}, focusing on  detail preservation, texture quality, and geometric accuracy. {Compared to existing methods  based on implicit fields or Gaussian splattings, our approach excels  in rendering detailed textures and producing geometry-aware representations. Besides, our field initialization demonstrates robust performance even when camera poses are inaccurate and multi-view consistency is not guaranteed, a common challenge in generated videos. We also compare our method with baselines using the same field initialization, and ours  delivers superior results.}

\textbf{Quantitative evaluation.} We provide the quantitative evaluation comparing our method with state-of-the-art works in Table~\ref{tab:main_results}. Metrics include Peak Signal-to-Noise Ratio (PSNR), Structural Similarity Index (SSIM), and Learned Perceptual Image Patch Similarity (LPIPS)~\cite{DBLP:conf/cvpr/ZhangIESW18}. Our method exhibits superiority over all baseline methods, \emph{e.g.}, $\sim$2.5 PSNR increase over SC-GS even with our field initialization for the averaged results.

We  collect 35 sub-videos from the SORA  webpage~\cite{videoworldsimulators2024}, including Drone\_Ancient\_Rome, Robot\_Scene, Seaside\_Aerial\_View, Mountain\_Horizontal\_View, Snow\_Sakura, etc. Comparison results are shown in Table~\ref{tab:sora_results}. Our full model achieves 4.24 PSNR improvement compared to the second-best method. Results prove the effectiveness of our method  on the unposed dynamic scenes.

\begin{figure*}[t]
    \centering
\includegraphics[width=1.8\columnwidth]{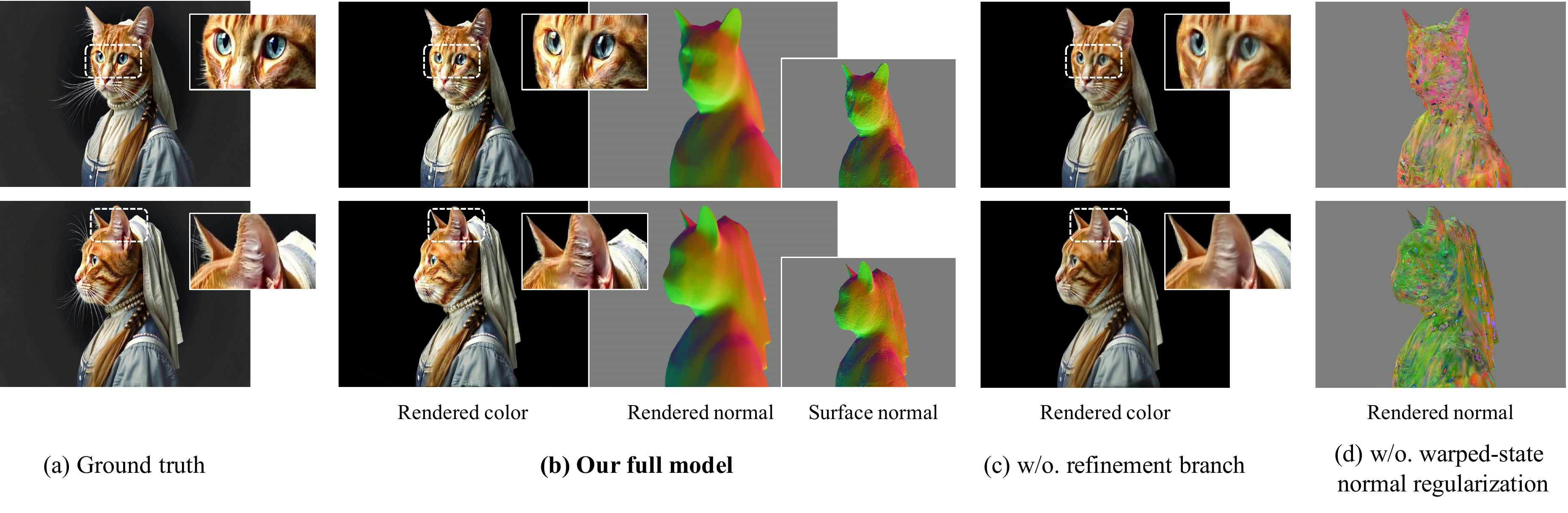}
    \vskip-0.1in
    \caption{{Ablation studies on our warped-state geometric regularization and dual-branch refinement strategy.} For our full model shown in (b), we provide our rendered color, rendered normal, and surface normal (estimated from the depth points for regularization). For comparison, we visualize the rendered color for the case without refinements in (c) and the rendered normal for the case without warped-state geometric regularization in (d), respectively. We showcase our model's fidelity with close-ups.}
    \label{fig:ablation}
\end{figure*}

\begin{figure}[t!]
    \centering\vskip-0.12in
    \includegraphics[width=0.9\columnwidth]{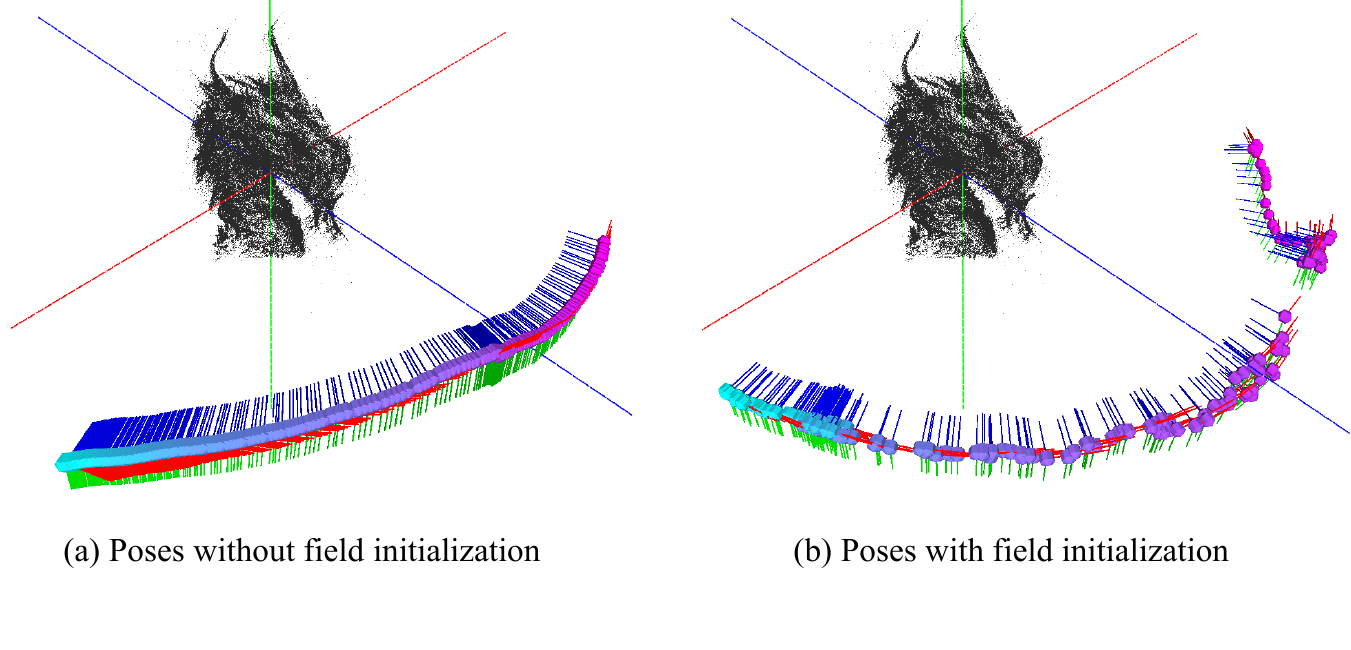}
        \vspace{-0.5 cm}
    \caption{Poses without and with applying  field initialization.} \vskip-0.09in
    \label{fig:supp-cam}
\end{figure}

\begin{figure}[t!]
   \centering \hskip-0.07in \vskip-0.1in
\includegraphics[width=1\columnwidth]{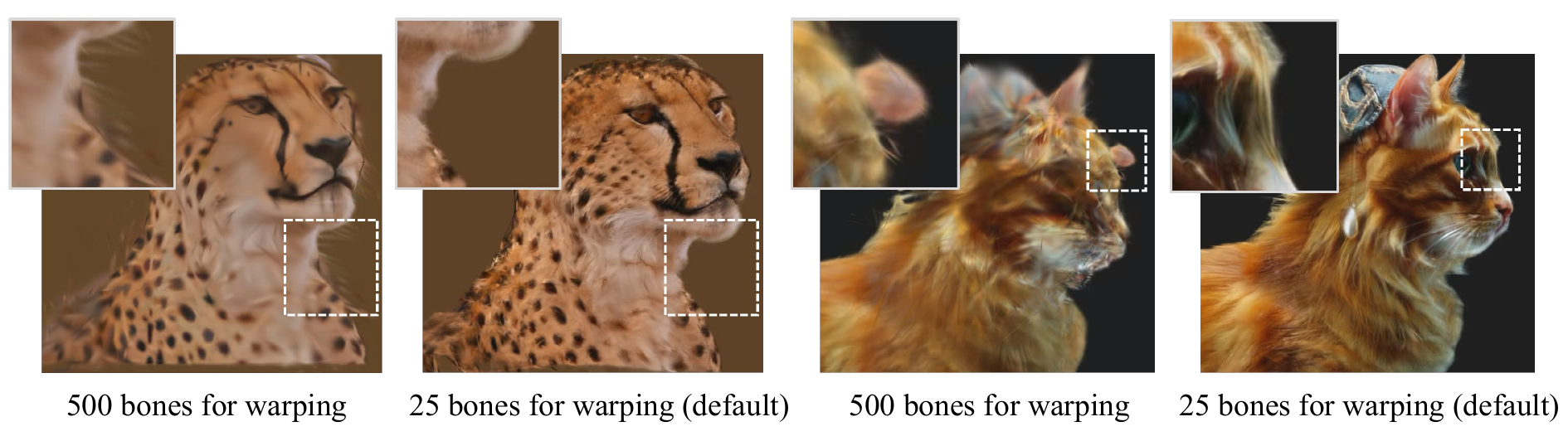}      \vskip-0.07in
\caption{The design of motion regularization alleviates over-fitting.}  \vskip-0.06in
   \label{fig:motion1}
\end{figure}

\begin{figure*}[t!]
    \centering   \vspace{-0.03in}
\includegraphics[width=1.6\columnwidth]{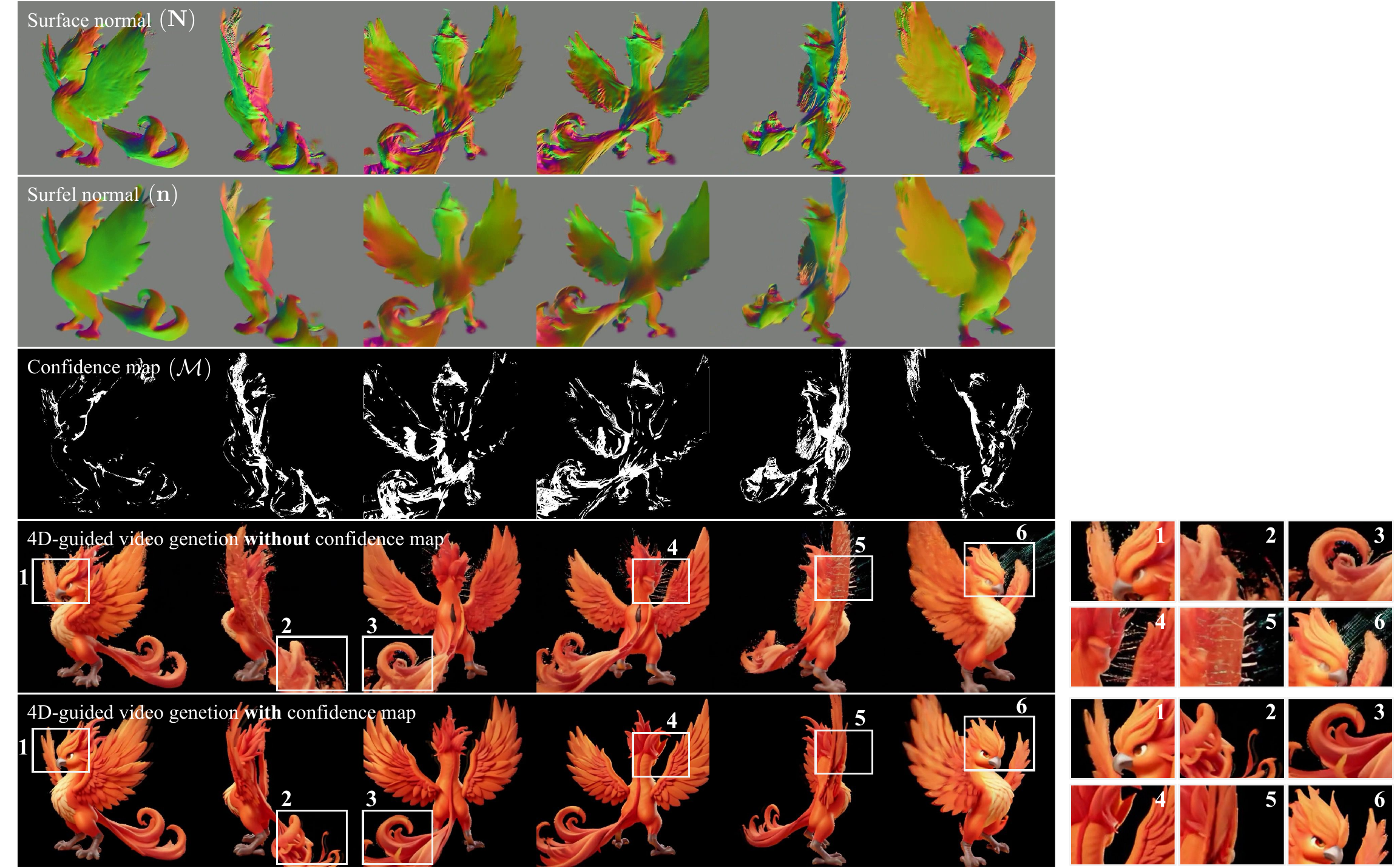}
    \caption{Ablation study on the 4D-guided video generation without or with using confidence map as the 4D filter.  We highlight regions to discern the differences in video results obtained with and without confidence map guidance.}
    \label{fig:compare-mask}   
\end{figure*}

\begin{figure}[t!]
   \centering
\includegraphics[width=0.8\columnwidth]{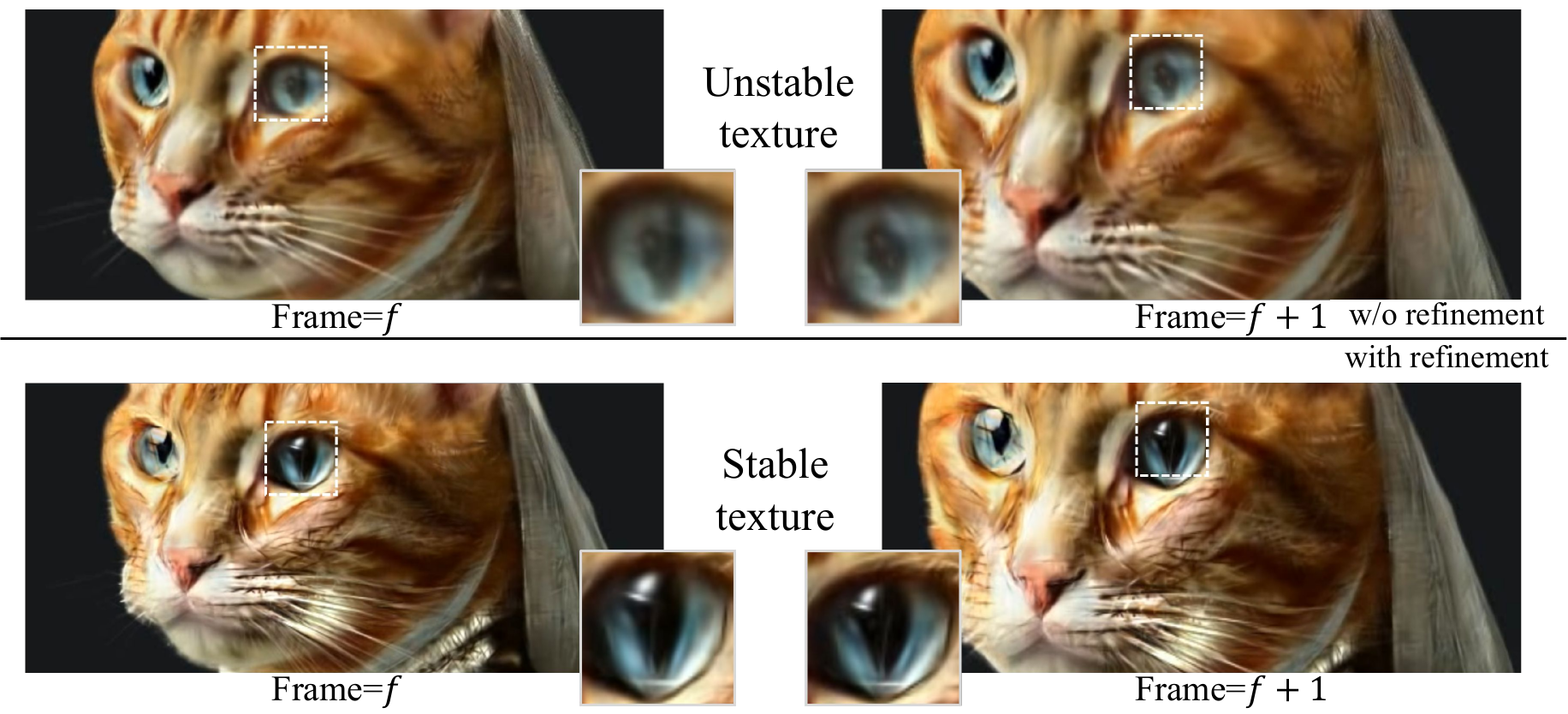}
   \vskip-0.06in
  \caption{Dual-branch refinement alleviates  flickering.}  \vspace{-0.03 cm}
   \label{fig:flick}
\end{figure}

\begin{figure}[t!]
\includegraphics[width=1\columnwidth]{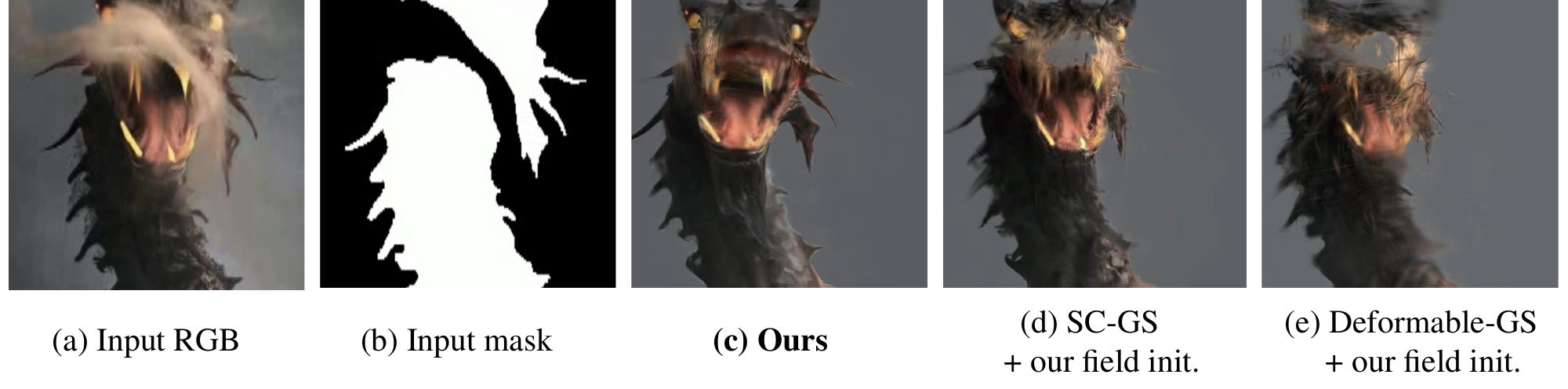}      \vskip-0.07in
\caption{Our reconstruction is robust to occlusion. Note that the input mask is obtained by  segmentation (not the    confidence map).} \vspace{-0.05in}
   \label{fig:motion2}
  \end{figure}

\begin{table*}[t]
\begin{minipage}{0.256\textwidth} 
\caption{{Quantitative results on \textbf{NeRF-DS} with/without GT poses.}}
\centering
\tablestyle{2pt}{0.92}
\resizebox{1\columnwidth}{!}{
\begin{tabular}{@{}l|ccc}
\toprule[1pt]
  \multirow{2}{*}{Method}  & \multicolumn{2}{c@{}}{  PSNR~$\uparrow$}\\ 
 & with GT poses & w/o GT poses\\
\midrule
HyperNeRF~\cite{DBLP:journals/tog/ParkSHBBGMS21} & 22.5 & 11.6\\
NeRF-DS~\cite{DBLP:conf/cvpr/Yan0L23} & 23.9  &12.1 \\
TiNeuVox-B~\cite{DBLP:conf/siggrapha/FangYWX00N022}& 21.5  &13.9\\
SC-GS~\cite{huang2023sc} &   24.1 &   14.6 \\
\textbf{Ours} & \best\textbf{24.3} & \best\textbf{19.0} \\
\bottomrule[1pt]
\end{tabular}}
\label{tab:quan-nerfds}
\end{minipage}\;
\begin{minipage}{0.72\textwidth} 
\centering
\caption{Quantitative ablation studies of the field initialization or  refinement.}
\tablestyle{0.8pt}{1.51}
\resizebox{1\columnwidth}{!}{
\begin{tabular}{@{}l|ccc|ccc|ccc|ccc@{}}
\toprule[1pt]
 \multirow{2}{*}{Method}& \multicolumn{3}{c@{}|}{Cat} & \multicolumn{3}{c@{}|}{Cheetah} & \multicolumn{3}{c@{}|}{Dragon}&  \multicolumn{3}{c@{}}{SORA benchmark}\\ 
& PSNR~$\uparrow$ & SSIM~$\uparrow$ & LPIPS~$\downarrow$ & PSNR~$\uparrow$ & 
SSIM~$\uparrow$ & LPIPS~$\downarrow$ & PSNR~$\uparrow$ & SSIM~$\uparrow$ & LPIPS~$\downarrow$& PSNR~$\uparrow$ & SSIM~$\uparrow$ & LPIPS~$\downarrow$ \\
\midrule
Ours w/o  field initialization  & 20.15 & 0.7961 & 0.2393 & 20.96 & 0.8194 & 0.1940 & 25.33 & 0.9146 & 0.0938 & 15.42 & 0.6167 & 0.2268 \\
Ours w/o dual branch refinement  & 24.19 & 0.8196 & 0.1797 & 24.10 & 0.8582 & 0.1242 & 27.71 & 0.9128 & 0.0687 & 18.57 & 0.6852  &  0.1945 \\
\textbf{Ours full} & \textbf{24.63} & \textbf{0.8432} & \textbf{0.1559} & \textbf{25.68} & \textbf{0.8843} & \textbf{0.1117} & \textbf{28.58} & \textbf{0.9392} & \textbf{0.0618} & \textbf{19.05} & \textbf{0.7323} & \textbf{0.1839}\\
\bottomrule[1pt]
\end{tabular}
}
\label{tab:supp-abl}
\end{minipage}
\end{table*}

\begin{figure}[t!]
    \centering
\includegraphics[width=0.9\columnwidth]{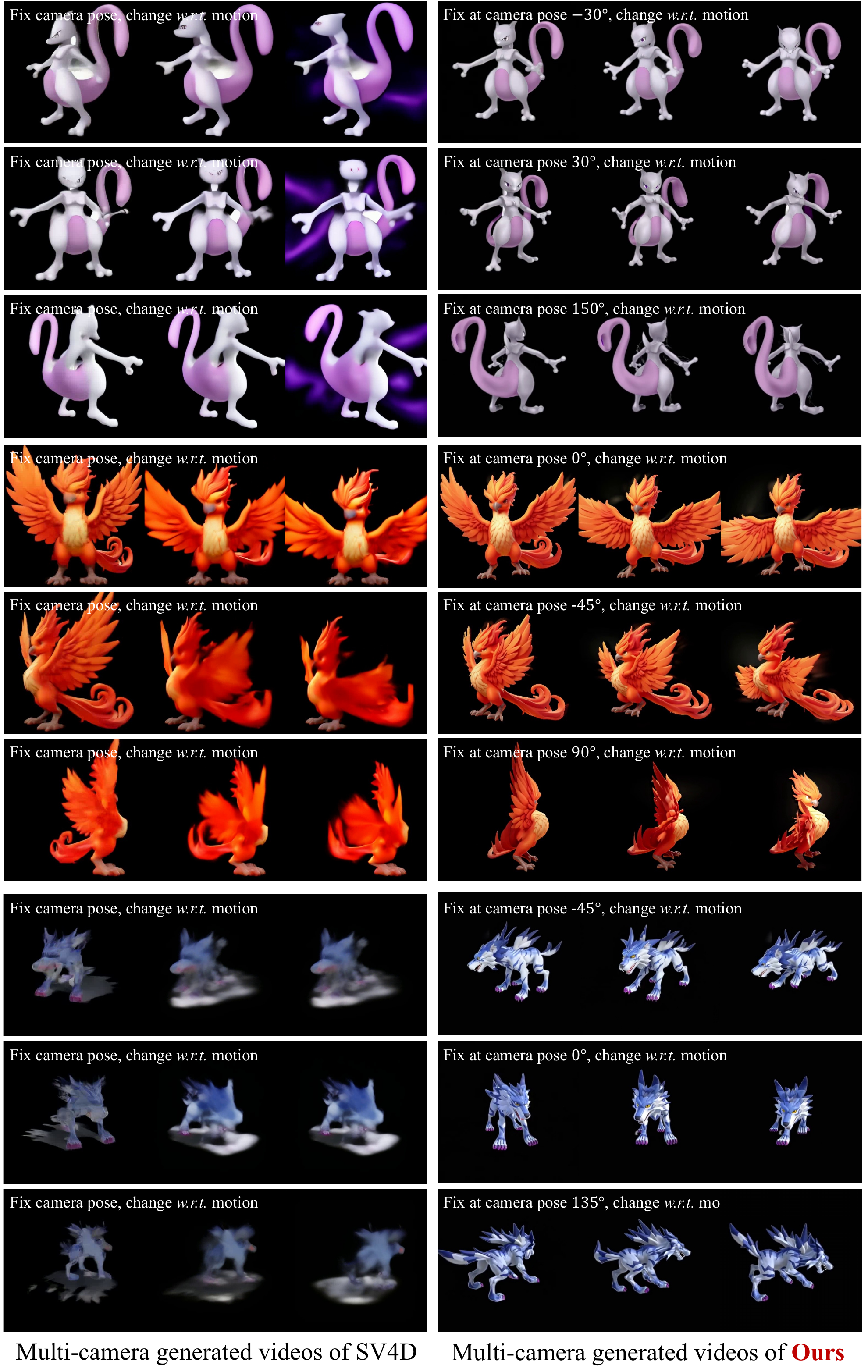}
\vspace{-0.2cm}
    \caption{Multi-camera video generation of our method compared with SV4D~\cite{DBLP:journals/corr/abs-2407-17470}. For either SV4D or ours, the three lines of each case follow the \textbf{same} motion sequence. } \vspace{-0.3in}
    \label{fig:compare-multicam}
\end{figure}

\subsection{Results of 4D Reconstruction from Realistic Videos}

To evaluate our DGS on realistic scenes for a comparison with other dynamic  methods, we choose  realistic scene-level benchmarks (Neural 3D Video dataset~\cite{DBLP:conf/cvpr/LiSZGL0SLGNL22}, NeRF-DS dataset~\cite{DBLP:conf/cvpr/Yan0L23}, and HyperNeRF dataset~\cite{DBLP:journals/tog/ParkSHBBGMS21}) and an object-level benchmark (D-NeRF dataset~\cite{pumarola2021d}). During the evaluation, we use PSNR, DSSIM, and LPIPS as evaluation metrics and follow the standard setting of state-of-the-art methods to perform training and evaluation. For the scene-level NeRF-DS data and the object-level D-NeRF data, we train the model for 80,000 iterations and start to perform geometric regularization on the 40,000$^\text{th}$ iteration. On D-NeRF and Neural 3D Video datasets, we perform additional groups of experiments when ground truth camera poses are unavailable. We train our field initialization stage (Fig.~\ref{fig:Video4DGen}) for 2,000 iterations which takes 10 minutes. For the scene-level Neural 3D Video data, we follow the standard setting to train  with 300 frames per scene at the resolution $1352\times1014$. We train the model by 30,000 iterations with  geometric regularization adding from the 10,000$^\text{th}$ iteration.

\textbf{Object-level 4D benchmark.} From the  experimental results on the D-NeRF dataset in Table~\ref{tab:dnerfwogt} (without GT poses), we observe that our DGS surpasses the second-best method by a large margin (29.06 vs. 20.05 for PSNR). When given GT poses, our method still outperforms existing methods as shown in Table~\ref{tab:dnerfgt}. We provide qualitative results in Fig.~\ref{fig:dnerf}, where we observe that the proposed method is especially superior at dynamic normals. 

\textbf{Scene-level (realistic) 4D benchmark.} From   results on  Neural 3D Video~\cite{DBLP:conf/cvpr/LiSZGL0SLGNL22}    in Table~\ref{tab:neural_3d_video}, our DGS achieves the best performance in capturing color and shows great superiority in modeling dynamic normals according to the visualizations in Fig.~\ref{fig:neural3dvideo}. Besides, in Table~\ref{tab:neural_3d_video}, we  provide comparisons of rendering latency on the benchmark, indicating the advantage of our rendering latency.

We also include comparisons on the NeRF-DS dataset~\cite{DBLP:conf/cvpr/Yan0L23} with methods including GS-IR~\cite{DBLP:conf/cvpr/LiangZFSJ24}, Gaussianshader~\cite{DBLP:conf/cvpr/JiangTLGLWM24}, etc, to evaluate the performance of methods when capturing changes in reflected color, as depicted in Fig.~\ref{fig:method_nerfds}. We compare a Quantitative results are provided in Table~\ref{tab:quan-nerfds} with or without (w/o) ground truth poses.  These results  indicate the effectiveness of our approach in handling reflected color variations. {Finally, we provide qualitative results on the HyperNeRF~\cite{DBLP:journals/tog/ParkSHBBGMS21}  dataset in Fig.~\ref{fig:method_hypernerf}, comparing with 
TiNeuVox~\cite{DBLP:conf/siggrapha/FangYWX00N022}, 4DGS~\cite{wu20234dgaussians}, Deformable-GS~\cite{yang2023deformable3dgs}, and Grid4D~\cite{xu2024grid4d}. Quantitative results are shown in the Appendix. Results highlight the superiority of our method in  maintaining both visual and geometric consistency during the warping process. }

\subsection{Results of Video Generation with 4D Guidance}

Besides high-fidelity 4D generation, our Video4DGen also demonstrates superior video generation with 4D guidance. We provide results early in Fig.~\ref{fig:intro-image-video}, where we introduce two novel video generation settings facilitated by 4D representation: multi-camera video generation and video generation that accommodates large pose and motion changes (\emph{e.g.}, up to 360$^\circ$). We also perform comparison with an existing method SV4D~\cite{DBLP:journals/corr/abs-2407-17470} as shown in Fig.~\ref{fig:compare-multicam}. By comparison, our results achieve much better performance in visual quality, multiview consistency, and temporal consistency.

\subsection{Ablation Studies and Discussions}
\label{subsec:ablation}
We conduct ablation studies to understand the contribution of each component in Video4DGen, especially DGS. We remove or alter specific elements of our model and observe the resulting performance changes in both appearance and geometry reconstruction.

\textbf{Geometric regularization.} We evaluate the impact of warped-state geometric regularization by disabling it during training. From Fig.~\ref{fig:ablation}(b)(d), we observe that when removing the regularization, there is a  degradation in the structural integrity of surface-aligned Gaussian surfels, leading to  inconsistency in  reconstructed  models.

\textbf{Field initialization.}
In Sec.~\ref{subsec:Video4DGen}, we propose field initialization to improve the optimization of poses and motion. We now verify the effectiveness of  field initialization by conducting comparison experiments to isolate the initialization. As shown in Table~\ref{tab:supp-abl}, we observe that when not applying field initialization, the performance of DGS  degrades. Poses without field initialization and with field initialization  are depicted in Fig.~\ref{fig:supp-cam}.

\textbf{Refinement strategy.} We examine the effectiveness of omitting refinement terms $\Delta\bR_k^{*}$ and $\Delta\bS_k^{*}$ during training, shown in Fig.~\ref{fig:ablation}(b)(c). The performance indicates that removing refinements increases the loss of fine-grained appearance details. We also find that in Fig.~\ref{fig:flick}, refinements are crucial for mitigating the texture flickering,  which accords with the analysis for Eq.~(\ref{eq:normal_large}).

\textbf{Motion regularization.} Our warping model uses a sparse-bone setting with 25 control bones, which performs better in motion regularization and reducing overfitting than increasing the number to 500, as shown in Fig.~\ref{fig:motion1}. Our motion regularization, which reduces Gaussian overfitting, is also robust against noise and occlusion which often occur in generated videos. In Fig.~\ref{fig:motion2}, the mask for the dragon's part hidden behind the sand is missing. Both SC-GS~\cite{huang2023sc} and Deformable-GS~\cite{yang2023deformable3dgs}   tend to overfit and  show a decline in performance. In contrast, ours is robust to occlusion.

\begin{figure*}[t!]
    \centering
    \includegraphics[width=1.8\columnwidth]{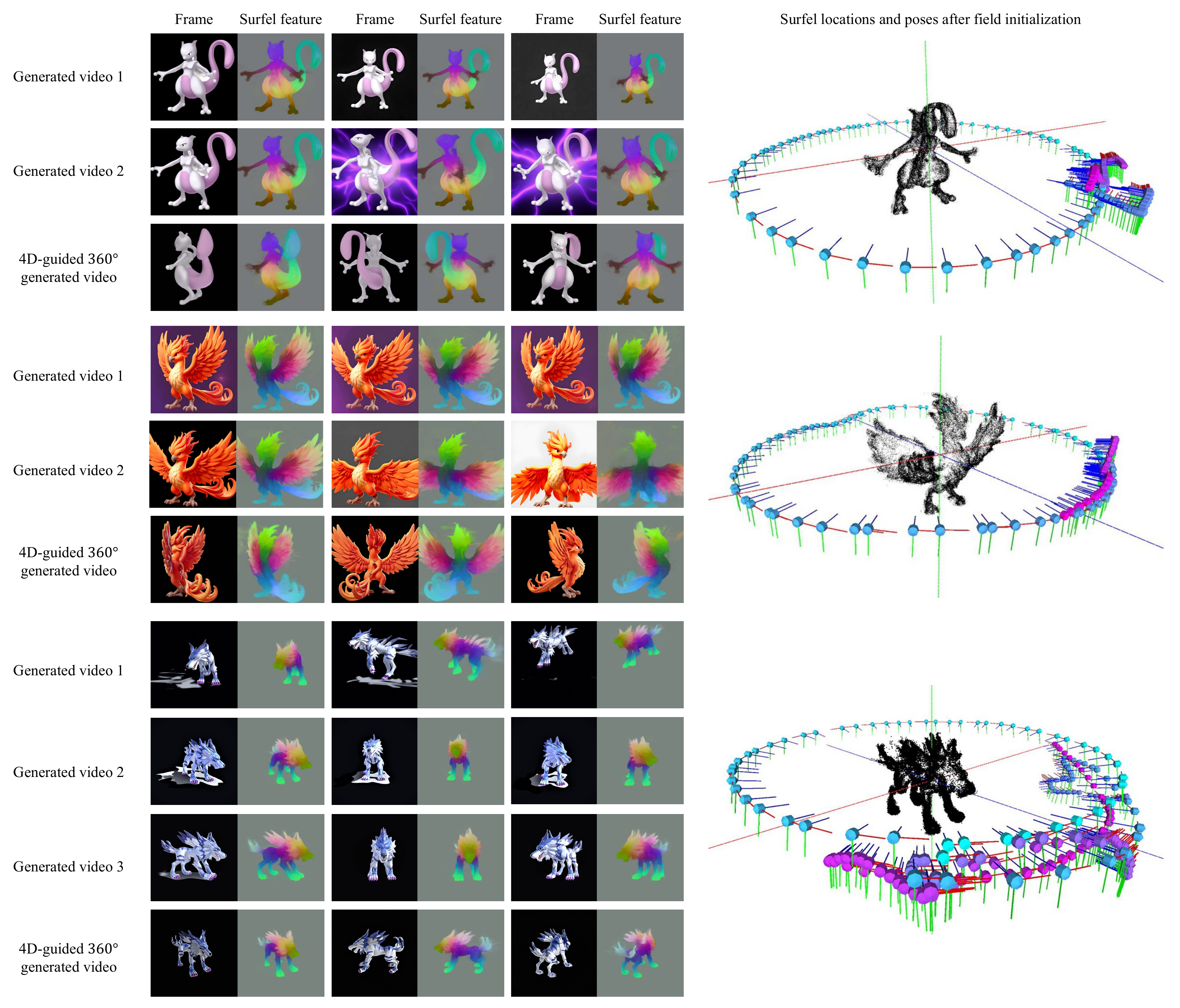}\vskip-0.09in
    \caption{Visualization of the multi-video DGS alignment (feature color) and the root poses optimized by field initialization. Besides, the 4D-guided 360$^\circ$ generated video enhances the 4D generation at a wide range of poses. }\vspace{0.2cm}
    \label{fig:poses_com}
\end{figure*}

\textbf{Confidence map for 4D-guided video generation.}  As previously detailed in Sec.~\ref{sec:vidgen}, to perform the 4D-guided video generation, we have devised a confidence map established by comparing the distance between the surfel normal  and the surface normal  to a predefined threshold. We provide  results of 4D-guided generated videos without or with the confidence map in Fig.~\ref{fig:compare-mask}. Upon comparison, it is evident that the result without the confidence map shows a higher degree of artifacts and noise.

\textbf{Multi-video DGS alignment, root poses, and 360$^\circ$ generated video.} We propose to align DGS for multiple generated videos and the root pose optimization in Sec.~\ref{sec:4dgen_multi}. Besides, in Sec.~\ref{sec:vidgen}, we  mention that the 4D-guided generated videos at novel views could be utilized to thereby enhance the 4D generation. In Fig.~\ref{fig:poses_com}, we provide visualization of these several parts.

\textbf{Additional ablation studies and discussions.} Please refer to the Appendix for more ablation studies and discussions: additional qualitative comparison, interpolation on time and view, mesh and depth extraction, reference videos and completing occluded views.

\section{Conclusion}
\label{sec:conclusion}

We present Video4DGen, a novel framework that jointly addresses two interrelated tasks: 4D generation from generated videos and 4D-guided video generation. We also propose Dynamic Gaussian Surfels (DGS) as a 4D representation in Video4DGen, to preserve high-fidelity appearance and geometry  during warping. Our experiments validate that Video4DGen outperforms existing methods in both quantitative metrics and qualitative evaluations, highlighting its superiority in generating realistic and immersive 4D content and 4D-guided video content. 

\textbf{Limitations and broader impact.} While Video4DGen with DGS presents a significant performance in 4D generation, currently there are still limitations such as the reliance on video quality, scalability challenges for large scenes, and computational difficulties in real-time applications. {We provide additional  cases in the Appendix to demonstrate potential challenges, constraints, or scenarios where the approach might underperform.} Additionally, when equipping Video4DGen with generative models, as with any generative technology, there is a risk of producing deceptive content which needs more caution.

\bibliographystyle{IEEEtran}
\balance
\bibliography{IEEEabrv,bare_jrnl_compsoc}

\begin{thebibliography}{100}
\providecommand{\url}[1]{#1}
\csname url@samestyle\endcsname
\providecommand{\newblock}{\relax}
\providecommand{\bibinfo}[2]{#2}
\providecommand{\BIBentrySTDinterwordspacing}{\spaceskip=0pt\relax}
\providecommand{\BIBentryALTinterwordstretchfactor}{4}
\providecommand{\BIBentryALTinterwordspacing}{\spaceskip=\fontdimen2\font plus
\BIBentryALTinterwordstretchfactor\fontdimen3\font minus
  \fontdimen4\font\relax}
\providecommand{\BIBforeignlanguage}[2]{{%
\expandafter\ifx\csname l@#1\endcsname\relax
\typeout{** WARNING: IEEEtran.bst: No hyphenation pattern has been}%
\typeout{** loaded for the language `#1'. Using the pattern for}%
\typeout{** the default language instead.}%
\else
\language=\csname l@#1\endcsname
\fi
#2}}
\providecommand{\BIBdecl}{\relax}
\BIBdecl

\bibitem{videoworldsimulators2024}
\BIBentryALTinterwordspacing
T.~Brooks, B.~Peebles, C.~Holmes, W.~DePue, Y.~Guo, L.~Jing, D.~Schnurr,
  J.~Taylor, T.~Luhman, E.~Luhman, C.~Ng, R.~Wang, and A.~Ramesh, ``Video
  generation models as world simulators,'' 2024. [Online]. Available:
  \url{https://openai.com/research/video-generation-models-as-world-simulators}
\BIBentrySTDinterwordspacing

\bibitem{bao2024vidu}
F.~Bao, C.~Xiang, G.~Yue, G.~He, H.~Zhu, K.~Zheng, M.~Zhao, S.~Liu, Y.~Wang,
  and J.~Zhu, ``Vidu: a highly consistent, dynamic and skilled text-to-video
  generator with diffusion models,'' \emph{arXiv preprint arXiv:2405.04233},
  2024.

\bibitem{chen2024v3d}
Z.~Chen, Y.~Wang, F.~Wang, Z.~Wang, and H.~Liu, ``V3d: Video diffusion models
  are effective 3d generators,'' \emph{arXiv preprint arXiv:2403.06738}, 2024.

\bibitem{voleti2024sv3d}
V.~Voleti, C.-H. Yao, M.~Boss, A.~Letts, D.~Pankratz, D.~Tochilkin, C.~Laforte,
  R.~Rombach, and V.~Jampani, ``Sv3d: Novel multi-view synthesis and 3d
  generation from a single image using latent video diffusion,'' \emph{arXiv
  preprint arXiv: 2403.12008}, 2024.

\bibitem{pumarola2021d}
A.~Pumarola, E.~Corona, G.~Pons-Moll, and F.~Moreno-Noguer, ``D-nerf: Neural
  radiance fields for dynamic scenes,'' in \emph{CVPR}, 2021.

\bibitem{TiNeuVox}
J.~Fang, T.~Yi, X.~Wang, L.~Xie, X.~Zhang, W.~Liu, M.~Nie\ss{}ner, and Q.~Tian,
  ``Fast dynamic radiance fields with time-aware neural voxels,'' in
  \emph{SIGGRAPH Asia}, 2022.

\bibitem{DBLP:journals/tog/ParkSHBBGMS21}
K.~Park, U.~Sinha, P.~Hedman, J.~T. Barron, S.~Bouaziz, D.~B. Goldman,
  R.~Martin{-}Brualla, and S.~M. Seitz, ``Hypernerf: a higher-dimensional
  representation for topologically varying neural radiance fields,''
  \emph{{ACM} Trans. Graph.}, 2021.

\bibitem{yang2023deformable3dgs}
Z.~Yang, X.~Gao, W.~Zhou, S.~Jiao, Y.~Zhang, and X.~Jin, ``Deformable 3d
  gaussians for high-fidelity monocular dynamic scene reconstruction,''
  \emph{arXiv preprint arXiv:2309.13101}, 2023.

\bibitem{wu20234dgaussians}
G.~Wu, T.~Yi, J.~Fang, L.~Xie, X.~Zhang, W.~Wei, W.~Liu, Q.~Tian, and
  W.~Xinggang, ``4d gaussian splatting for real-time dynamic scene rendering,''
  \emph{arXiv preprint arXiv:2310.08528}, 2023.

\bibitem{huang2023sc}
Y.-H. Huang, Y.-T. Sun, Z.~Yang, X.~Lyu, Y.-P. Cao, and X.~Qi, ``Sc-gs:
  Sparse-controlled gaussian splatting for editable dynamic scenes,''
  \emph{arXiv preprint arXiv:2312.14937}, 2023.

\bibitem{DBLP:conf/cvpr/LiCLX24}
Z.~Li, Z.~Chen, Z.~Li, and Y.~Xu, ``Spacetime gaussian feature splatting for
  real-time dynamic view synthesis,'' in \emph{CVPR}, 2024.

\bibitem{Huang2DGS2024}
B.~Huang, Z.~Yu, A.~Chen, A.~Geiger, and S.~Gao, ``2d gaussian splatting for
  geometrically accurate radiance fields,'' in \emph{SIGGRAPH}.\hskip 1em plus
  0.5em minus 0.4em\relax Association for Computing Machinery, 2024.

\bibitem{Dai2024GaussianSurfels}
P.~Dai, J.~Xu, W.~Xie, X.~Liu, H.~Wang, and W.~Xu, ``High-quality surface
  reconstruction using gaussian surfels,'' in \emph{SIGGRAPH}, 2024.

\bibitem{schonberger2016structure}
J.~L. Schonberger and J.-M. Frahm, ``Structure-from-motion revisited,'' in
  \emph{CVPR}, 2016.

\bibitem{DBLP:conf/eccv/SchonbergerZFP16}
J.~L. Sch{\"{o}}nberger, E.~Zheng, J.~Frahm, and M.~Pollefeys, ``Pixelwise view
  selection for unstructured multi-view stereo,'' in \emph{ECCV}, 2016.

\bibitem{buehler2001unstructured}
C.~Buehler, M.~Bosse, L.~McMillan, S.~Gortler, and M.~Cohen, ``Unstructured
  lumigraph rendering,'' in \emph{Proceedings of the 28th annual conference on
  Computer graphics and interactive techniques}, 2001.

\bibitem{debevec1996modeling}
P.~E. Debevec, C.~J. Taylor, and J.~Malik, ``Modeling and rendering
  architecture from photographs: A hybrid geometry-and image-based approach,''
  in \emph{Proceedings of the 23rd annual conference on Computer graphics and
  interactive techniques}, 1996.

\bibitem{waechter2014let}
M.~Waechter, N.~Moehrle, and M.~Goesele, ``Let there be color! large-scale
  texturing of 3d reconstructions,'' in \emph{ECCV}, 2014.

\bibitem{wood2000surface}
D.~N. Wood, D.~I. Azuma, K.~Aldinger, B.~Curless, T.~Duchamp, D.~H. Salesin,
  and W.~Stuetzle, ``Surface light fields for 3d photography,'' in
  \emph{Proceedings of the 27th annual conference on Computer graphics and
  interactive techniques}, 2000.

\bibitem{riegler2020free}
G.~Riegler and V.~Koltun, ``Free view synthesis,'' in \emph{ECCV}, 2020.

\bibitem{thies2019deferred}
J.~Thies, M.~Zollh{\"o}fer, and M.~Nie{\ss}ner, ``Deferred neural rendering:
  Image synthesis using neural textures,'' \emph{{ACM} Trans. Graph.}, 2019.

\bibitem{sitzmann2019deepvoxels}
V.~Sitzmann, J.~Thies, F.~Heide, M.~Nie{\ss}ner, G.~Wetzstein, and
  M.~Zollhofer, ``Deepvoxels: Learning persistent 3d feature embeddings,'' in
  \emph{CVPR}, 2019.

\bibitem{lombardi2019neural}
S.~Lombardi, T.~Simon, J.~Saragih, G.~Schwartz, A.~Lehrmann, and Y.~Sheikh,
  ``Neural volumes: Learning dynamic renderable volumes from images,''
  \emph{arXiv preprint arXiv:1906.07751}, 2019.

\bibitem{penner2017soft}
E.~Penner and L.~Zhang, ``Soft 3d reconstruction for view synthesis,''
  \emph{{ACM} Trans. Graph.}, 2017.

\bibitem{kutulakos2000theory}
K.~N. Kutulakos and S.~M. Seitz, ``A theory of shape by space carving,''
  \emph{IJCV}, 2000.

\bibitem{zhou2018stereo}
T.~Zhou, R.~Tucker, J.~Flynn, G.~Fyffe, and N.~Snavely, ``Stereo magnification:
  Learning view synthesis using multiplane images,'' \emph{arXiv preprint
  arXiv:1805.09817}, 2018.

\bibitem{flynn2019deepview}
J.~Flynn, M.~Broxton, P.~Debevec, M.~DuVall, G.~Fyffe, R.~Overbeck, N.~Snavely,
  and R.~Tucker, ``Deepview: View synthesis with learned gradient descent,'' in
  \emph{CVPR}, 2019.

\bibitem{mildenhall2019local}
B.~Mildenhall, P.~P. Srinivasan, R.~Ortiz-Cayon, N.~K. Kalantari,
  R.~Ramamoorthi, R.~Ng, and A.~Kar, ``Local light field fusion: Practical view
  synthesis with prescriptive sampling guidelines,'' \emph{{ACM} Trans.
  Graph.}, 2019.

\bibitem{srinivasan2019pushing}
P.~P. Srinivasan, R.~Tucker, J.~T. Barron, R.~Ramamoorthi, R.~Ng, and
  N.~Snavely, ``Pushing the boundaries of view extrapolation with multiplane
  images,'' in \emph{CVPR}, 2019.

\bibitem{srinivasan2020lighthouse}
P.~P. Srinivasan, B.~Mildenhall, M.~Tancik, J.~T. Barron, R.~Tucker, and
  N.~Snavely, ``Lighthouse: Predicting lighting volumes for spatially-coherent
  illumination,'' in \emph{CVPR}, 2020.

\bibitem{tucker2020single}
R.~Tucker and N.~Snavely, ``Single-view view synthesis with multiplane
  images,'' in \emph{CVPR}, 2020.

\bibitem{DBLP:conf/eccv/MildenhallSTBRN20}
B.~Mildenhall, P.~P. Srinivasan, M.~Tancik, J.~T. Barron, R.~Ramamoorthi, and
  R.~Ng, ``Nerf: Representing scenes as neural radiance fields for view
  synthesis,'' in \emph{ECCV}, 2020.

\bibitem{barron2021mip}
J.~T. Barron, B.~Mildenhall, M.~Tancik, P.~Hedman, R.~Martin-Brualla, and P.~P.
  Srinivasan, ``Mip-nerf: A multiscale representation for anti-aliasing neural
  radiance fields,'' in \emph{ICCV}, 2021.

\bibitem{barron2023zipnerf}
J.~T. Barron, B.~Mildenhall, D.~Verbin, P.~P. Srinivasan, and P.~Hedman,
  ``Zip-nerf: Anti-aliased grid-based neural radiance fields,'' \emph{arXiv
  preprint arXiv:2304.06706}, 2023.

\bibitem{kerbl3Dgaussians}
B.~Kerbl, G.~Kopanas, T.~Leimk{\"u}hler, and G.~Drettakis, ``3d gaussian
  splatting for real-time radiance field rendering,'' \emph{{ACM} Trans.
  Graph.}, 2023.

\bibitem{ma2021deblur}
L.~Ma, X.~Li, J.~Liao, Q.~Zhang, X.~Wang, J.~Wang, and P.~V. Sander,
  ``Deblur-nerf: Neural radiance fields from blurry images,'' \emph{arXiv
  preprint arXiv:2111.14292}, 2021.

\bibitem{wang20224k}
Z.~Wang, L.~Li, Z.~Shen, L.~Shen, and L.~Bo, ``4k-nerf: High fidelity neural
  radiance fields at ultra high resolutions,'' \emph{arXiv preprint
  arXiv:2212.04701}, 2022.

\bibitem{reiser2023merf}
C.~Reiser, R.~Szeliski, D.~Verbin, P.~P. Srinivasan, B.~Mildenhall, A.~Geiger,
  J.~T. Barron, and P.~Hedman, ``Merf: Memory-efficient radiance fields for
  real-time view synthesis in unbounded scenes,'' \emph{arXiv preprint
  arXiv:2302.12249}, 2023.

\bibitem{hedman2021snerg}
P.~Hedman, P.~P. Srinivasan, B.~Mildenhall, J.~T. Barron, and P.~Debevec,
  ``Baking neural radiance fields for real-time view synthesis,'' \emph{ICCV},
  2021.

\bibitem{yu2021plenoctrees}
A.~Yu, R.~Li, M.~Tancik, H.~Li, R.~Ng, and A.~Kanazawa, ``Plenoctrees for
  real-time rendering of neural radiance fields,'' in \emph{CVPR}, 2021.

\bibitem{reiser2021kilonerf}
C.~Reiser, S.~Peng, Y.~Liao, and A.~Geiger, ``Kilonerf: Speeding up neural
  radiance fields with thousands of tiny mlps,'' in \emph{CVPR}, 2021.

\bibitem{liu2020neural}
L.~Liu, J.~Gu, K.~Zaw~Lin, T.-S. Chua, and C.~Theobalt, ``Neural sparse voxel
  fields,'' in \emph{NeurIPS}, 2020.

\bibitem{rebain2021derf}
D.~Rebain, W.~Jiang, S.~Yazdani, K.~Li, K.~M. Yi, and A.~Tagliasacchi, ``Derf:
  Decomposed radiance fields,'' in \emph{CVPR}, 2021.

\bibitem{lindell2021autoint}
D.~B. Lindell, J.~N. Martel, and G.~Wetzstein, ``Autoint: Automatic integration
  for fast neural volume rendering,'' in \emph{CVPR}, 2021.

\bibitem{garbin2021fastnerf}
S.~J. Garbin, M.~Kowalski, M.~Johnson, J.~Shotton, and J.~Valentin, ``Fastnerf:
  High-fidelity neural rendering at 200fps,'' in \emph{ICCV}, 2021.

\bibitem{Hu_2022_CVPR}
T.~Hu, S.~Liu, Y.~Chen, T.~Shen, and J.~Jia, ``Efficientnerf efficient neural
  radiance fields,'' in \emph{CVPR}, 2022.

\bibitem{cao2022mobiler2l}
J.~Cao, H.~Wang, P.~Chemerys, V.~Shakhrai, J.~Hu, Y.~Fu, D.~Makoviichuk,
  S.~Tulyakov, and J.~Ren, ``Real-time neural light field on mobile devices,''
  \emph{arXiv preprint arXiv:2212.08057}, 2022.

\bibitem{wang2022r2l}
H.~Wang, J.~Ren, Z.~Huang, K.~Olszewski, M.~Chai, Y.~Fu, and S.~Tulyakov,
  ``R2l: Distilling neural radiance field to neural light field for efficient
  novel view synthesis,'' in \emph{ECCV}, 2022.

\bibitem{lombardi2021mixture}
S.~Lombardi, T.~Simon, G.~Schwartz, M.~Zollhoefer, Y.~Sheikh, and J.~Saragih,
  ``Mixture of volumetric primitives for efficient neural rendering,''
  \emph{arXiv preprint arXiv:2103.01954}, 2021.

\bibitem{guedon2023sugar}
A.~Gu{\'e}don and V.~Lepetit, ``Sugar: Surface-aligned gaussian splatting for
  efficient 3d mesh reconstruction and high-quality mesh rendering,''
  \emph{arXiv preprint arXiv:2311.12775}, 2023.

\bibitem{chen2023text}
Z.~Chen, F.~Wang, Y.~Wang, and H.~Liu, ``Text-to-3d using gaussian splatting,''
  in \emph{CVPR}, 2024.

\bibitem{tang2024lgm}
J.~Tang, Z.~Chen, X.~Chen, T.~Wang, G.~Zeng, and Z.~Liu, ``Lgm: Large
  multi-view gaussian model for high-resolution 3d content creation,''
  \emph{arXiv preprint arXiv:2402.05054}, 2024.

\bibitem{xu2024grm}
Y.~Xu, Z.~Shi, W.~Yifan, S.~Peng, C.~Yang, Y.~Shen, and W.~Gordon, ``Grm: Large
  gaussian reconstruction model for efficient 3d reconstruction and
  generation,'' \emph{arXiv preprint arXiv: 2403.14621}, 2024.

\bibitem{liu2024citygaussian}
Y.~Liu, H.~Guan, C.~Luo, L.~Fan, J.~Peng, and Z.~Zhang, ``Citygaussian:
  Real-time high-quality large-scale scene rendering with gaussians,''
  \emph{arXiv preprint arXiv: 2404.01133}, 2024.

\bibitem{shuai2024LoG}
Q.~Shuai, H.~Guo, Z.~Xu, H.~Lin, S.~Peng, H.~Bao, and X.~Zhou, ``Real-time view
  synthesis for large scenes with millions of square meters,'' 2024.

\bibitem{hierarchicalgaussians24}
B.~Kerbl, A.~Meuleman, G.~Kopanas, M.~Wimmer, A.~Lanvin, and G.~Drettakis, ``A
  hierarchical 3d gaussian representation for real-time rendering of very large
  datasets,'' \emph{{ACM} Trans. Graph.}, 2024.

\bibitem{li2022tava}
R.~Li, J.~Tanke, M.~Vo, M.~Zollhofer, J.~Gall, A.~Kanazawa, and C.~Lassner,
  ``Tava: Template-free animatable volumetric actors,'' 2022.

\bibitem{peng2021neural}
S.~Peng, Y.~Zhang, Y.~Xu, Q.~Wang, Q.~Shuai, H.~Bao, and X.~Zhou, ``Neural
  body: Implicit neural representations with structured latent codes for novel
  view synthesis of dynamic humans,'' in \emph{CVPR}, 2021.

\bibitem{su2021anerf}
S.-Y. Su, F.~Yu, M.~Zollh{\"o}fer, and H.~Rhodin, ``A-nerf: Articulated neural
  radiance fields for learning human shape, appearance, and pose,'' in
  \emph{NeurIPS}, 2021.

\bibitem{zhang2021stnerf}
Z.~Jiakai, L.~Xinhang, Y.~Xinyi, Z.~Fuqiang, Z.~Yanshun, W.~Minye,
  Z.~Yingliang, X.~Lan, and Y.~Jingyi, ``Editable free-viewpoint video using a
  layered neural representation,'' in \emph{SIGGRAPH}, 2021.

\bibitem{wang2023animatabledreamer}
X.~Wang, Y.~Wang, J.~Ye, Z.~Wang, F.~Sun, P.~Liu, L.~Wang, K.~Sun, X.~Wang, and
  B.~He, ``Animatabledreamer: Text-guided non-rigid 3d model generation and
  reconstruction with canonical score distillation,'' \emph{arXiv preprint
  arXiv:2312.03795}, 2023.

\bibitem{DBLP:conf/nips/0011SW0T22}
L.~Li, Z.~Shen, Z.~Wang, L.~Shen, and P.~Tan, ``Streaming radiance fields for
  3d video synthesis,'' in \emph{NeurIPS}, 2022.

\bibitem{wang2023rpd}
Y.~Wang, Y.~Dong, F.~Sun, and X.~Yang, ``Root pose decomposition towards
  generic non-rigid 3d reconstruction with monocular videos,'' in \emph{ICCV},
  2023.

\bibitem{attal2023hyperreel}
B.~Attal, J.-B. Huang, C.~Richardt, M.~Zollhoefer, J.~Kopf, M.~O'Toole, and
  C.~Kim, ``Hyperreel: High-fidelity 6-dof video with ray-conditioned
  sampling,'' \emph{arXiv preprint arXiv:2301.02238}, 2023.

\bibitem{song2022nerfplayer}
L.~Song, A.~Chen, Z.~Li, Z.~Chen, L.~Chen, J.~Yuan, Y.~Xu, and A.~Geiger,
  ``Nerfplayer: A streamable dynamic scene representation with decomposed
  neural radiance fields,'' \emph{arXiv preprint arXiv:2210.15947}, 2022.

\bibitem{cao2023hexplane}
A.~Cao and J.~Johnson, ``Hexplane: a fast representation for dynamic scenes,''
  \emph{arXiv preprint arXiv:2301.09632}, 2023.

\bibitem{wang2022mixed}
F.~Wang, S.~Tan, X.~Li, Z.~Tian, and H.~Liu, ``Mixed neural voxels for fast
  multi-view video synthesis,'' \emph{arXiv preprint arXiv:2212.00190}, 2022.

\bibitem{wang2022fourier}
L.~Wang, J.~Zhang, X.~Liu, F.~Zhao, Y.~Zhang, Y.~Zhang, M.~Wu, J.~Yu, and
  L.~Xu, ``Fourier plenoctrees for dynamic radiance field rendering in
  real-time,'' in \emph{CVPR}, 2022.

\bibitem{bansal20204d}
A.~Bansal, M.~Vo, Y.~Sheikh, D.~Ramanan, and S.~Narasimhan, ``4d visualization
  of dynamic events from unconstrained multi-view videos,'' in \emph{CVPR},
  2020.

\bibitem{peng2023representing}
S.~Peng, Y.~Yan, Q.~Shuai, H.~Bao, and X.~Zhou, ``Representing volumetric
  videos as dynamic mlp maps,'' in \emph{CVPR}, 2023.

\bibitem{wang2023masked}
F.~Wang, Z.~Chen, G.~Wang, Y.~Song, and H.~Liu, ``Masked space-time hash
  encoding for efficient dynamic scene reconstruction,'' in \emph{NeurIPS},
  2023.

\bibitem{DBLP:conf/cvpr/LiSZGL0SLGNL22}
T.~Li, M.~Slavcheva, M.~Zollh{\"{o}}fer, S.~Green, C.~Lassner, C.~Kim,
  T.~Schmidt, S.~Lovegrove, M.~Goesele, R.~A. Newcombe, and Z.~Lv, ``Neural 3d
  video synthesis from multi-view video,'' in \emph{CVPR}, 2022.

\bibitem{kplanes_2023}
{Sara Fridovich-Keil and Giacomo Meanti}, F.~R. Warburg, B.~Recht, and
  A.~Kanazawa, ``K-planes: Explicit radiance fields in space, time, and
  appearance,'' in \emph{CVPR}, 2023.

\bibitem{DBLP:conf/cvpr/YangVNRVJ22}
G.~Yang, M.~Vo, N.~Neverova, D.~Ramanan, A.~Vedaldi, and H.~Joo, ``Banmo:
  Building animatable 3d neural models from many casual videos,'' in
  \emph{CVPR}, 2022.

\bibitem{dycheck}
H.~Gao, R.~Li, S.~Tulsiani, B.~Russell, and A.~Kanazawa, ``Dynamic novel-view
  synthesis: A reality check,'' in \emph{NeurIPS}, 2022.

\bibitem{DBLP:conf/iccv/ParkSBBGSM21}
K.~Park, U.~Sinha, J.~T. Barron, S.~Bouaziz, D.~B. Goldman, S.~M. Seitz, and
  R.~Martin{-}Brualla, ``Nerfies: Deformable neural radiance fields,'' in
  \emph{ICCV}, 2021.

\bibitem{shao2023tensor4d}
R.~Shao, Z.~Zheng, H.~Tu, B.~Liu, H.~Zhang, and Y.~Liu, ``Tensor4d: Efficient
  neural 4d decomposition for high-fidelity dynamic reconstruction and
  rendering,'' in \emph{CVPR}, 2023.

\bibitem{liu2023robust}
Y.-L. Liu, C.~Gao, A.~Meuleman, H.-Y. Tseng, A.~Saraf, C.~Kim, Y.-Y. Chuang,
  J.~Kopf, and J.-B. Huang, ``Robust dynamic radiance fields,'' in \emph{CVPR},
  2023.

\bibitem{liu2022devrf}
J.-W. Liu, Y.-P. Cao, W.~Mao, W.~Zhang, D.~J. Zhang, J.~Keppo, Y.~Shan, X.~Qie,
  and M.~Z. Shou, ``Devrf: Fast deformable voxel radiance fields for dynamic
  scenes,'' \emph{arXiv preprint arXiv:2205.15723}, 2022.

\bibitem{Zhao_2022_CVPR}
F.~Zhao, W.~Yang, J.~Zhang, P.~Lin, Y.~Zhang, J.~Yu, and L.~Xu, ``Humannerf:
  Efficiently generated human radiance field from sparse inputs,'' in
  \emph{CVPR}, 2022.

\bibitem{tretschk2021non}
E.~Tretschk, A.~Tewari, V.~Golyanik, M.~Zollh{\"o}fer, C.~Lassner, and
  C.~Theobalt, ``Non-rigid neural radiance fields: Reconstruction and novel
  view synthesis of a dynamic scene from monocular video,'' in \emph{ICCV},
  2021.

\bibitem{jiang2022alignerf}
Y.~Jiang, P.~Hedman, B.~Mildenhall, D.~Xu, J.~T. Barron, Z.~Wang, and T.~Xue,
  ``Alignerf: High-fidelity neural radiance fields via alignment-aware
  training,'' \emph{arXiv preprint arXiv:2211.09682}, 2022.

\bibitem{du2021nerflow}
Y.~Du, Y.~Zhang, H.-X. Yu, J.~B. Tenenbaum, and J.~Wu, ``Neural radiance flow
  for 4d view synthesis and video processing,'' in \emph{ICCV}, 2021.

\bibitem{gao2021dynamic}
C.~Gao, A.~Saraf, J.~Kopf, and J.-B. Huang, ``Dynamic view synthesis from
  dynamic monocular video,'' in \emph{ICCV}, 2021.

\bibitem{li2020neural}
Z.~Li, S.~Niklaus, N.~Snavely, and O.~Wang, ``Neural scene flow fields for
  space-time view synthesis of dynamic scenes,'' in \emph{CVPR}, 2021.

\bibitem{xian2021space}
W.~Xian, J.-B. Huang, J.~Kopf, and C.~Kim, ``Space-time neural irradiance
  fields for free-viewpoint video,'' in \emph{CVPR}, 2021.

\bibitem{luiten2023dynamic}
J.~Luiten, G.~Kopanas, B.~Leibe, and D.~Ramanan, ``Dynamic 3d gaussians:
  Tracking by persistent dynamic view synthesis,'' in \emph{3DV}, 2024.

\bibitem{DBLP:conf/siggraph/DuanWDHCC24}
Y.~Duan, F.~Wei, Q.~Dai, Y.~He, W.~Chen, and B.~Chen, ``4d-rotor gaussian
  splatting: Towards efficient novel view synthesis for dynamic scenes,'' in
  \emph{SIGGRAPH}, 2024.

\bibitem{poole2022dreamfusion}
B.~Poole, A.~Jain, J.~T. Barron, and B.~Mildenhall, ``Dreamfusion: Text-to-3d
  using 2d diffusion,'' \emph{arXiv preprint arXiv:2209.14988}, 2022.

\bibitem{wang2023prolificdreamer}
Z.~Wang, C.~Lu, Y.~Wang, F.~Bao, C.~Li, H.~Su, and J.~Zhu, ``Prolificdreamer:
  High-fidelity and diverse text-to-3d generation with variational score
  distillation,'' in \emph{NeurIPS}, 2023.

\bibitem{lin2023magic3d}
C.-H. Lin, J.~Gao, L.~Tang, T.~Takikawa, X.~Zeng, X.~Huang, K.~Kreis,
  S.~Fidler, M.-Y. Liu, and T.-Y. Lin, ``Magic3d: High-resolution text-to-3d
  content creation,'' in \emph{CVPR}, 2023.

\bibitem{chen2023fantasia3d}
R.~Chen, Y.~Chen, N.~Jiao, and K.~Jia, ``Fantasia3d: Disentangling geometry and
  appearance for high-quality text-to-3d content creation,'' in \emph{ICCV},
  2023.

\bibitem{singer2023text}
U.~Singer, S.~Sheynin, A.~Polyak, O.~Ashual, I.~Makarov, F.~Kokkinos, N.~Goyal,
  A.~Vedaldi, D.~Parikh, J.~Johnson \emph{et~al.}, ``Text-to-4d dynamic scene
  generation,'' \emph{arXiv preprint arXiv:2301.11280}, 2023.

\bibitem{ling2023align}
H.~Ling, S.~W. Kim, A.~Torralba, S.~Fidler, and K.~Kreis, ``Align your
  gaussians: Text-to-4d with dynamic 3d gaussians and composed diffusion
  models,'' \emph{arXiv preprint arXiv:2312.13763}, 2023.

\bibitem{bah20244dfy}
S.~Bahmani, I.~Skorokhodov, V.~Rong, G.~Wetzstein, L.~Guibas, P.~Wonka,
  S.~Tulyakov, J.~J. Park, A.~Tagliasacchi, and D.~B. Lindell, ``4d-fy:
  Text-to-4d generation using hybrid score distillation sampling,'' in
  \emph{CVPR)}, 2024.

\bibitem{hong2023lrm}
Y.~Hong, K.~Zhang, J.~Gu, S.~Bi, Y.~Zhou, D.~Liu, F.~Liu, K.~Sunkavalli,
  T.~Bui, and H.~Tan, ``Lrm: Large reconstruction model for single image to
  3d,'' \emph{arXiv preprint arXiv:2311.04400}, 2023.

\bibitem{zou2023triplane}
Z.-X. Zou, Z.~Yu, Y.-C. Guo, Y.~Li, D.~Liang, Y.-P. Cao, and S.-H. Zhang,
  ``Triplane meets gaussian splatting: Fast and generalizable single-view 3d
  reconstruction with transformers,'' \emph{arXiv preprint arXiv:2312.09147},
  2023.

\bibitem{wang2024crm}
Z.~Wang, Y.~Wang, Y.~Chen, C.~Xiang, S.~Chen, D.~Yu, C.~Li, H.~Su, and J.~Zhu,
  ``Crm: Single image to 3d textured mesh with convolutional reconstruction
  model,'' \emph{arXiv preprint arXiv:2403.05034}, 2024.

\bibitem{long2023wonder3d}
X.~Long, Y.-C. Guo, C.~Lin, Y.~Liu, Z.~Dou, L.~Liu, Y.~Ma, S.-H. Zhang,
  M.~Habermann, C.~Theobalt \emph{et~al.}, ``Wonder3d: Single image to 3d using
  cross-domain diffusion,'' \emph{arXiv preprint arXiv:2310.15008}, 2023.

\bibitem{lu2023direct2}
Y.~Lu, J.~Zhang, S.~Li, T.~Fang, D.~McKinnon, Y.~Tsin, L.~Quan, X.~Cao, and
  Y.~Yao, ``Direct2. 5: Diverse text-to-3d generation via multi-view 2.5 d
  diffusion,'' \emph{arXiv preprint arXiv:2311.15980}, 2023.

\bibitem{DBLP:journals/corr/abs-2407-12781}
S.~Bahmani, I.~Skorokhodov, A.~Siarohin, W.~Menapace, G.~Qian, M.~Vasilkovsky,
  H.~Lee, C.~Wang, J.~Zou, A.~Tagliasacchi, D.~B. Lindell, and S.~Tulyakov,
  ``{VD3D:} taming large video diffusion transformers for 3d camera control,''
  \emph{CoRR}, vol. abs/2407.12781, 2024.

\bibitem{DBLP:journals/corr/abs-2406-02509}
D.~Xu, W.~Nie, C.~Liu, S.~Liu, J.~Kautz, Z.~Wang, and A.~Vahdat, ``Camco:
  Camera-controllable 3d-consistent image-to-video generation,'' \emph{CoRR},
  vol. abs/2406.02509, 2024.

\bibitem{DBLP:journals/corr/abs-2405-17414}
Z.~Kuang, S.~Cai, H.~He, Y.~Xu, H.~Li, L.~J. Guibas, and G.~Wetzstein,
  ``Collaborative video diffusion: Consistent multi-video generation with
  camera control,'' \emph{CoRR}, vol. abs/2405.17414, 2024.

\bibitem{DBLP:journals/corr/abs-2404-02101}
H.~He, Y.~Xu, Y.~Guo, G.~Wetzstein, B.~Dai, H.~Li, and C.~Yang, ``Cameractrl:
  Enabling camera control for text-to-video generation,'' 2024.

\bibitem{DBLP:conf/cvpr/CaiCGHWW24}
S.~Cai, D.~Ceylan, M.~Gadelha, C.~P. Huang, T.~Y. Wang, and G.~Wetzstein,
  ``Generative rendering: Controllable 4d-guided video generation with 2d
  diffusion models,'' in \emph{CVPR}, 2024.

\bibitem{DBLP:conf/nips/SongE19}
Y.~Song and S.~Ermon, ``Generative modeling by estimating gradients of the data
  distribution,'' in \emph{NeurIPS}, 2019.

\bibitem{DBLP:conf/cvpr/YangSJVCCRFL21}
G.~Yang, D.~Sun, V.~Jampani, D.~Vlasic, F.~Cole, H.~Chang, D.~Ramanan, W.~T.
  Freeman, and C.~Liu, ``{LASR:} learning articulated shape reconstruction from
  a monocular video,'' in \emph{ICCV}, 2021.

\bibitem{DBLP:conf/si3d/KavanCZO07}
L.~Kavan, S.~Collins, J.~Z{\'{a}}ra, and C.~O'Sullivan, ``Skinning with dual
  quaternions,'' in \emph{SI3D}, 2007.

\bibitem{botsch2005high}
M.~Botsch, A.~Hornung, M.~Zwicker, and L.~Kobbelt, ``High-quality surface
  splatting on today's gpus,'' in \emph{Proceedings Eurographics/IEEE VGTC
  Symposium Point-Based Graphics}, 2005.

\bibitem{DBLP:conf/nips/WangLLTKW21}
P.~Wang, L.~Liu, Y.~Liu, C.~Theobalt, T.~Komura, and W.~Wang, ``Neus: Learning
  neural implicit surfaces by volume rendering for multi-view reconstruction,''
  in \emph{NeurIPS}, 2021.

\bibitem{DBLP:journals/corr/abs-2206-15258}
H.~Cai, W.~Feng, X.~Feng, Y.~Wang, and J.~Zhang, ``Neural surface
  reconstruction of dynamic scenes with monocular {RGB-D} camera,'' in
  \emph{NeurIPS}, 2022.

\bibitem{oquab2023dinov2}
M.~Oquab, T.~Darcet, T.~Moutakanni, H.~Vo, M.~Szafraniec, V.~Khalidov,
  P.~Fernandez, D.~Haziza, F.~Massa, A.~El-Nouby \emph{et~al.}, ``Dinov2:
  Learning robust visual features without supervision,'' \emph{arXiv preprint
  arXiv:2304.07193}, 2023.

\bibitem{DBLP:conf/iclr/LiuG023}
X.~Liu, C.~Gong, and Q.~Liu, ``Flow straight and fast: Learning to generate and
  transfer data with rectified flow,'' in \emph{ICLR}, 2023.

\bibitem{DBLP:conf/nips/KarrasAAL22}
T.~Karras, M.~Aittala, T.~Aila, and S.~Laine, ``Elucidating the design space of
  diffusion-based generative models,'' in \emph{NeurIPS}, 2022.

\bibitem{DBLP:conf/iclr/0011SKKEP21}
Y.~Song, J.~Sohl{-}Dickstein, D.~P. Kingma, A.~Kumar, S.~Ermon, and B.~Poole,
  ``Score-based generative modeling through stochastic differential
  equations,'' in \emph{ICLR}, 2021.

\bibitem{DBLP:conf/siggrapha/FangYWX00N022}
J.~Fang, T.~Yi, X.~Wang, L.~Xie, X.~Zhang, W.~Liu, M.~Nie{\ss}ner, and Q.~Tian,
  ``Fast dynamic radiance fields with time-aware neural voxels,'' in
  \emph{SIGGRAPH}, 2022.

\bibitem{DBLP:conf/cvpr/ShaoZTL0L23}
R.~Shao, Z.~Zheng, H.~Tu, B.~Liu, H.~Zhang, and Y.~Liu, ``Tensor4d: Efficient
  neural 4d decomposition for high-fidelity dynamic reconstruction and
  rendering,'' in \emph{CVPR}, 2023.

\bibitem{DBLP:conf/cvpr/Fridovich-KeilM23}
S.~Fridovich{-}Keil, G.~Meanti, F.~R. Warburg, B.~Recht, and A.~Kanazawa,
  ``K-planes: Explicit radiance fields in space, time, and appearance,'' in
  \emph{CVPR}, 2023.

\bibitem{DBLP:conf/iccv/GuoSDCY0DZ023}
X.~Guo, J.~Sun, Y.~Dai, G.~Chen, X.~Ye, X.~Tan, E.~Ding, Y.~Zhang, and J.~Wang,
  ``Forward flow for novel view synthesis of dynamic scenes,'' in \emph{ICCV},
  2023.

\bibitem{DBLP:journals/tvcg/SongCLCCYXG23}
L.~Song, A.~Chen, Z.~Li, Z.~Chen, L.~Chen, J.~Yuan, Y.~Xu, and A.~Geiger,
  ``Nerfplayer: {A} streamable dynamic scene representation with decomposed
  neural radiance fields,'' \emph{{IEEE} Trans. Vis. Comput. Graph.}, 2023.

\bibitem{DBLP:conf/cvpr/Attal0RZ0OK23}
B.~Attal, J.~Huang, C.~Richardt, M.~Zollh{\"{o}}fer, J.~Kopf, M.~O'Toole, and
  C.~Kim, ``Hyperreel: High-fidelity 6-dof video with ray-conditioned
  sampling,'' in \emph{CVPR}, 2023.

\bibitem{DBLP:conf/iccv/WangTLTS023}
F.~Wang, S.~Tan, X.~Li, Z.~Tian, Y.~Song, and H.~Liu, ``Mixed neural voxels for
  fast multi-view video synthesis,'' in \emph{ICCV}, 2023.

\bibitem{DBLP:conf/iclr/YangYP024}
Z.~Yang, H.~Yang, Z.~Pan, and L.~Zhang, ``Real-time photorealistic dynamic
  scene representation and rendering with 4d gaussian splatting,'' in
  \emph{ICLR}, 2024.

\bibitem{DBLP:conf/nips/YarivKMGABL20}
L.~Yariv, Y.~Kasten, D.~Moran, M.~Galun, M.~Atzmon, R.~Basri, and Y.~Lipman,
  ``Multiview neural surface reconstruction by disentangling geometry and
  appearance,'' in \emph{NeurIPS}, 2020.

\bibitem{DBLP:conf/cvpr/ZhangIESW18}
R.~Zhang, P.~Isola, A.~A. Efros, E.~Shechtman, and O.~Wang, ``The unreasonable
  effectiveness of deep features as a perceptual metric,'' in \emph{CVPR},
  2018.

\bibitem{DBLP:conf/cvpr/Yan0L23}
Z.~Yan, C.~Li, and G.~H. Lee, ``Nerf-ds: Neural radiance fields for dynamic
  specular objects,'' in \emph{CVPR}, 2023.

\bibitem{DBLP:journals/corr/abs-2407-17470}
Y.~Xie, C.~Yao, V.~Voleti, H.~Jiang, and V.~Jampani, ``{SV4D:} dynamic 3d
  content generation with multi-frame and multi-view consistency,''
  \emph{CoRR}, vol. abs/2407.17470, 2024.

\bibitem{DBLP:conf/cvpr/LiangZFSJ24}
Z.~Liang, Q.~Zhang, Y.~Feng, Y.~Shan, and K.~Jia, ``{GS-IR:} 3d gaussian
  splatting for inverse rendering,'' in \emph{CVPR}, 2024.

\bibitem{DBLP:conf/cvpr/JiangTLGLWM24}
Y.~Jiang, J.~Tu, Y.~Liu, X.~Gao, X.~Long, W.~Wang, and Y.~Ma, ``Gaussianshader:
  3d gaussian splatting with shading functions for reflective surfaces,'' in
  \emph{CVPR}, 2024.

\bibitem{xu2024grid4d}
X.~Jiawei, F.~Zexin, Y.~Jian, and X.~Jin, ``{Grid4D}: {4D} decomposed hash
  encoding for high-fidelity dynamic scene rendering,'' in \emph{NeurIPS},
  2024.

\end{thebibliography}

\end{document}